\documentclass[preprint,12pt]{elsarticle}

\journal{Theoretical Computer Science}

\usepackage{pstricks,pst-node,pst-plot,pst-coil,pst-text,multido,pst-3d,pst-tree,pst-grad,subfigure,colortbl,color,array}
\usepackage{graphicx,longtable,amsmath,amscd,amsthm,pspicture,epsfig,bm,pifont,xspace,graphics,graphicx,amssymb,epsf,citesort,wrapfig,ctable,color,url,floatflt}

\newtheorem{Theorem}{Theorem}

\newtheorem{Lemma}{Lemma}

\newtheorem{Definition}{Definition}

\newtheorem{Proposition}{Proposition}
\newtheorem{fact}{Fact}
\newtheorem{remark}{Remark}

\newenvironment{proof-sketch}{{\noindent\bf Sketch of proof.\ }}{\hfill{\Pisymbol{pzd}{113}}\vspace{0.1in}}

\newcommand{\TB}{\vspace*{-0.1ex}}\newcommand{\TiE}{\setlength{\itemsep}{-1ex}}

\newcommand{\IE}{{\em i.e.}\xspace}
\newcommand{\EG}{{\em e.g.}\xspace}
\newcommand{\EA}{{\em et al.}\xspace}

\newcommand{\tx}{^{\rm th}}

\newcommand{\cS}{\mathcal{S}}
\newcommand{\Q}{{\mathcal{Q}}}
\newcommand{\cSn}{{\mathcal{S}}_{\mathrm{non-trivial}}}

\newcommand{\cU}{\mathcal{U}}

\newcommand{\cD}{\mathcal{D}}
\newcommand{\UD}{\mathsf{D}_u}

\newcommand{\ignore}[1]{}

\newcommand{\be}[1]{\begin{equation}\label{#1}}
\newcommand{\ee}{\end{equation}}
\newcommand{\bi}{\begin{itemize}}
\newcommand{\ei}{\end{itemize}}
\newcommand{\ben}{\begin{enumerate}}
\newcommand{\een}{\end{enumerate}}

\newcommand{\BSP}{{\sc Bsp}}
\newcommand{\PAR}{{\sc Par}}

\newcommand{\x}{\mathbf{x}}
\newcommand{\y}{\mathbf{y}}

\newcommand{\aw}{{\alpha_{\mathrm{\scriptstyle worst}}}}

\newcommand{\comment}[1]{}

\newcommand{\vol}{\mathsf{Volume}}
\newcommand{\party}{\mathsf{party}}
\newcommand{\rr}{\mathsf{r}}
\newcommand{\nbr}{\mathsf{Nbr}}

\begin{document}

\begin{frontmatter}



\title{On Communication Protocols that Compute Almost Privately\tnoteref{label1}}
\tnotetext[label1]{A preliminary version of this paper appeared in the $4\tx$ Symposium on Algorithmic Game Theory, 
G. Persiano (Ed.), LNCS 6982, Springer-Verlag, 44-56, 2011.}


\author[a1]{Marco Comi}
\ead{ingmarco85@gmail.com}
\author[a1]{Bhaskar DasGupta\corref{a3}}
\ead{bdasgup@uic.edu}
\ead[url]{http://www.cs.uic.edu/~dasgupta}
\author[a2]{Michael Schapira}
\ead{schapiram@huji.ac.il}
\author[a1]{Venkatakumar Srinivasan}
\ead{vsrini7@uic.edu}
\address[a1]{Department of Computer Science, University of Illinois at Chicago, IL 60607} 
\address[a2]{School of Computer Science and Engineering, The Hebrew University of Jerusalem, 91904 Jerusalem, Israel}
\cortext[a3]{Corresponding author. Phone: 312-355-1319, Fax: 312-413-0024.}

\begin{abstract}
We further investigate and generalize the approximate privacy model recently introduced by Feigenbaum \EA~\cite{FJS10}. We explore the privacy properties of a natural
class of communication protocols that we refer to as ``{\em dissection protocols}''. Informally, in a dissection protocol
the communicating parties are restricted to answering questions of the form ``{\em Is your input between the values $\alpha$ and $\beta$}
(under a pre-defined order over the possible inputs)?''. We prove that for a large class
of functions, called {\em tiling functions}, there {\em always} exists a dissection protocol that provides a {\em constant
average-case privacy approximation ratio} for uniform or ``almost uniform'' probability distributions over inputs. 
To establish this result we present an
interesting connection between the approximate privacy framework and
basic concepts in computational geometry. We show that such a good privacy approximation ratio for tiling functions
does {\em not}, in general, exist in the {\em worst case}.
We also discuss extensions of the basic setup to more than two parties and to non-tiling functions,
and provide calculations of privacy approximation ratios for two functions of interest.
\end{abstract}

\end{frontmatter}



\section{Introduction}

Consider the following interaction between two parties, Alice and Bob.
Each of the two parties, Alice and Bob, holds a {\em private} input, $x_{\mathrm{bob}}$ and $y_{\mathrm{alice}}$ respectively, not known to the other party. The two parties aim to compute a function $f$ of the two private inputs.
Alice and Bob alternately query each other to make available
a {\em small} amount of information about their private inputs, \EG, an answer to a range query on their private inputs or
a few bits of their private inputs. This process ends when each of them has seen enough information to be able to
compute the value of $f(x_{\mathrm{bob}},y_{\mathrm{alice}})$.
The central question that is the focus of this paper is:

\begin{quotation}
\noindent
{\em Can we design a communication protocol
whose execution reveals, to both Alice and Bob, as well as to any eavesdropper, as little information as possible
about the other's private input beyond what is necessary to compute the function value}?
\end{quotation}

\noindent
Note that there are two conflicting constraints: Alice and Bob need to communicate
sufficient information for computing the function value, but would prefer not to communicate too much
information about their private inputs.
This setting can be generalized in an obvious manner to
$d>1$ parties $\party_1,\party_2,\dots,\party_d$ computing a $d$-ary $f$ by querying the parties in round-robin order,
allowing each party to broadcast information about its private input (via a public communication channel).

Privacy preserving computational models such as the one described above have become an important research area due to the
increasingly widespread usage of sensitive data in networked environments, as evidenced by distributed computing applications, game-theoretic
settings (\EG, auctions) and more.
Over the years computer scientists have explored many {\em quantifications} of privacy in computation. Much
of this research focused on designing {\em perfectly} privacy-preserving protocols, \IE, protocols whose execution
reveals {\em no} information about the parties' private inputs beyond that implied by the outcome of the computation. Unfortunately,
perfect privacy is often either {\em impossible}, or {\em infeasibly costly} to achieve (\EG, requiring impractically extensive communication steps).
To overcome this, researchers have also investigated various notions of {\em approximate privacy}~\cite{D06,FJS10}.

In this paper, we adopt the approximate privacy framework of~\cite{FJS10} that quantifies
approximate privacy via the {\em privacy approximation ratios} ({\sc Par}s) of
protocols for computing a deterministic function of two private inputs.
{\em Informally}, \PAR\ captures the objective that an observer of the transcript of the entire protocol
will not be able to distinguish the real inputs of the two communicating parties from {\em as large a set as possible} of other inputs.
To capture this intuition, \cite{FJS10} makes use of the machinery of communication-complexity theory to provide a geometric and combinatorial
interpretation of protocols. \cite{FJS10} formulates both the worst-case and the average-case version of {\sc Par}s and studies the tradeoff between privacy
preservation and communication complexity for several functions of interest.

\subsection{Motivations from Mechanism Designs} 

An original motivation of this line of research, as explained in details in~\cite{FJS10}, comes from privacy concerns in auction theory in Economics.
A traditionally desired goal of designing auction mechanisms is to ensure that it is {\em incentive compatible}, \IE, bidders fare best by letting their truthful bids 
known. However, more recently, another {\em complementary} goal that has gained significant attention, specially in the context of online auctions, 
is to preserve privacy of the bidders, \IE, bidders reveal as little information as necessary to auctioneers for optimal outcomes.
To give an example, consider a $2^{\mathrm{nd}}$-price Vickrey auction of an item via a straightforward protocol in which the price of the item
is incrementally increased until the winner is determined. However, the protocol reveals more information than what is {\em absolutely necessary}, namely 
the information about the identity of the winner (with revealing his/her bid) together with the bid of the second-highest bidder, and revealing 
such additional information could put the winner at a disadvantage in the bidding process of a similar item in the future since the auctioneer
could set a lower reserve price. In this paper, we consider a generalized version of the setting that captures applications of the above type as well as other applications in 
multi-party computation.

\section{Summary of Our Contributions}
\label{contribution}

Any investigation of approximate privacy for multi-party computation starts by defining how we quantify
approximate privacy.
In this paper, we use the combinatorial framework of~\cite{FJS10} for quantification of approximate privacy
for two parties via \PAR s and present its natural extension to three or more parties.
Often, parties' inputs have a natural ordering, \EG, the private input of a party belongs to some range of integers $\{L,L+1,\dots,M\}$
(as is the case when computing, say, the maximum or minimum of two inputs). When designing protocols for such environments,
a natural restriction is to only allow the protocol to ask each party questions of the form
``{\sf\em Is your input between the values $\alpha$ and $\beta$ (under this natural order over possible inputs)?}''. We refer to this type of protocols as {\em dissection protocols} and study the privacy properties of this natural class of protocols. We note that the bisection and $c$-bisection protocols for the millionaires problem and other problems in~\cite{FJS10}, as well as the bisection auction in~\cite{muller,muller2},
all fall within this category of protocols. Our findings are summarized below.

\vspace{0.1in}
\noindent
{\bf Average- and worst-case {\sc Par}s for tiling functions for two party computation.}
We first consider a broad class of functions,  referred to as the {\em tiling functions} in the sequel,
that encompasses several well-studied functions (\EG, Vickrey's second-price auctions).
Informally, a two-variable tiling function is a function whose output space can be
viewed as a collection of disjoint combinatorial rectangles in the two-dimensional plane, where the function has the same value
within each rectangle. A first natural question for investigation is to
classify those tiling functions for which there exists a perfectly privacy-preserving dissection protocol.
We observe that for every Boolean tiling functions (\IE, tiling functions which output binary values) {\em this is indeed the case}.
In contrast, for tiling functions with a range of just three values, perfectly privacy-preserving computation is no longer necessarily possible
(even when not restricted to dissection protocols).

We next turn our attention to {\sc Par}s. We prove that for {\em every} tiling function there exists a dissection protocol that achieves a
constant \PAR\ in the average case (that is, when the parties' private values are drawn from an uniform or {\em almost} uniform probability distribution).
To establish this result, we make use of results on the binary space partitioning problems studied in the computational geometry literature.
We complement this positive result for dissection protocols with the following negative result:
{\em there exist tiling functions for which no dissection protocol can achieve a constant \PAR\ in the worst-case}.

\vspace{0.1in}
\noindent
{\bf Extensions to non-tiling functions and three-party communication.}
We discuss two extensions of the above results. We explain how our constant average-case \PAR\ result for tiling functions can be extended to a family of
``almost'' tiling functions. In addition, we consider the case of {\em more than two} parties. We show that in this setting
it is {\em no longer true} that for every tiling function there exists a dissection protocol that achieves a constant \PAR\ in the average case.
Namely, we exhibit a three-dimensional tiling function for which {\em every} dissection protocol exhibits {\em exponential} average- and worst-case \PAR s,
{\em even when an unlimited number of communication steps is allowed}.

\vspace{0.1in}
\noindent
{\bf PARs for the {\sf set covering} and {\sf equality} functions.}
\cite{FJS10} presents bounds on the average-case and the worst-case {\sc Par}s of the bisection protocol --- a special case of dissection protocols --- for
several functions (Yao's millionaires' problem, Vickrey's second-price auction, and others). We analyze the {\sc Par}s of the bisection protocol for
two well-studied Boolean functions: the {\sf set-covering} and {\sf equality} functions;
the {\sf equality} function provides a useful testbed for evaluating privacy preserving protocols~\cite{BCKO93}~\cite[Example $1.21$]{HN97}
and set-covering type of functions are useful for studying the differences between deterministic and non-deterministic
communication complexities~\cite[Section $10.4$]{HN97}.
We show that, for both functions, the bisection protocol {\em fails to achieve} good {\sc Par}s in both the average-case and the worst-case.

\section{Summary of Prior Related Works}

\subsection{Privacy-preserving Computation}

Privacy-preserving computation has been the subject of extensive research and has been approached from
information-theoretic~\cite{BCKO93}, cryptographic~\cite{CCD88},
statistical~\cite{KL11}, communication complexity~\cite{K92,Y79}, statistical database query~\cite{D06} and other perspectives~\cite{HN97}.
Among these, most relevant to our work is the approximate privacy framework of Feigenbaum \EA~\cite{FJS10} that presents
a metric for quantifying privacy preservation building on the work of Chor and Kushilevitz~\cite{CK91} on characterizing perfectly privately
computable computation and on the work of Kushilevitz~\cite{K92} on the communication complexity of perfectly private computation.
The bisection, $c$-bisection and bounded bisection protocols of~\cite{FJS10} fall within our category of dissection protocol since
we allow the input space of each party to be divided into two subsets of arbitrary size.
There are also some other formulations of perfectly and approximately privacy-preserving computation in the literature, but they
are inapplicable in our context.
For example, the differential privacy model (see~\cite{D06}) approaches privacy in a different context
via adding noise to the result of a database query in such a way as to preserve the privacy of the
individual database records but still have the result
convey nontrivial information,

\subsection{Binary space partition (\BSP)}

{\sc Bsp}s present a way to implement a {\em geometric divide-and-conquer} strategy and is an extremely popular
approach in numerous applications such as hidden surface removal, ray-tracing, visibility problems, solid geometry,
motion planning and spatial databases (\EG, see~\cite{T05}). However, to the best of our knowledge,
a connection between {\sc Bsp}s bounds such as in~\cite{paterson90,paterson92,BDM02,dAF92} and approximate privacy has not been explored before.

\section{The Model and Basic Definitions}

\subsection{Two-party Approximate Privacy Model of~\cite{FJS10}}
\label{setup1}

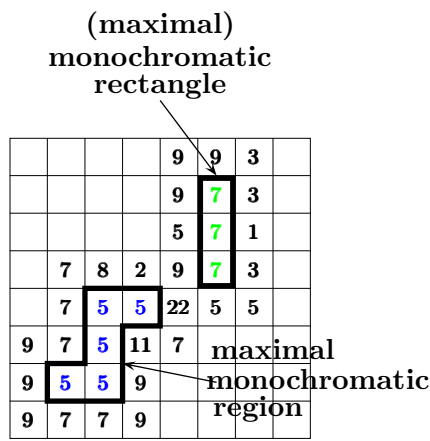
\begin{wrapfigure}[16]{R}{2.4in}
\begin{pspicture}(-0.2,0)(3.8,5.5)
\psset{xunit=1cm,yunit=1cm}
\psframe[linewidth=0.2pt,linecolor=black,fillstyle=none,fillcolor=lightgray,origin={0,0}](0,0)(4,4)
\psline[linewidth=0.2pt](1,0)(1,4)
\psline[linewidth=0.2pt,origin={-0.5,0}](1,0)(1,4)
\psline[linewidth=0.2pt](2,0)(2,4)
\psline[linewidth=0.2pt,origin={-0.5,0}](2,0)(2,4)
\psline[linewidth=0.2pt](3,0)(3,4)
\psline[linewidth=0.2pt,origin={-0.5,0}](3,0)(3,4)
\psline[linewidth=0.2pt,origin={-0.5,0}](4,0)(4,4)
\psline[linewidth=0.2pt](0,1)(4,1)
\psline[linewidth=0.2pt,origin={0,-0.5}](0,1)(4,1)
\psline[linewidth=0.2pt](0,2)(4,2)
\psline[linewidth=0.2pt,origin={0,-0.5}](0,2)(4,2)
\psline[linewidth=0.2pt](0,3)(4,3)
\psline[linewidth=0.2pt,origin={0,-0.5}](0,3)(4,3)
\psline[linewidth=0.2pt,origin={0,-0.5}](0,4)(4,4)
\pspolygon[linewidth=2pt,linecolor=black,fillstyle=none,fillcolor=lightgray](0.5,0.5)(1.5,0.5)(1.5,1.5)(2,1.5)(2,2)(1,2)(1,1.5)(1,1)(0.5,1)
\rput(0.75,0.75){\color{blue}$\pmb{\scriptstyle 5}$}
\rput(1.25,0.75){\color{blue}$\pmb{\scriptstyle 5}$}
\rput(1.25,1.25){\color{blue}$\pmb{\scriptstyle 5}$}
\rput(1.25,1.75){\color{blue}$\pmb{\scriptstyle 5}$}
\rput(1.75,1.75){\color{blue}$\pmb{\scriptstyle 5}$}
\rput(0.75,0.25){$\pmb{\scriptstyle 7}$}
\rput(1.25,0.25){$\pmb{\scriptstyle 7}$}
\rput(1.75,0.25){$\pmb{\scriptstyle 9}$}
\rput(1.75,0.75){$\pmb{\scriptstyle 9}$}
\rput(1.75,1.25){$\pmb{\scriptstyle 11}$}
\rput(2.25,1.25){$\pmb{\scriptstyle 7}$}
\rput(2.25,1.75){$\pmb{\scriptstyle 22}$}
\rput(2.25,2.25){$\pmb{\scriptstyle 9}$}
\rput(1.75,2.25){$\pmb{\scriptstyle 2}$}
\rput(1.25,2.25){$\pmb{\scriptstyle 8}$}
\rput(0.75,2.25){$\pmb{\scriptstyle 7}$}
\rput(0.75,1.75){$\pmb{\scriptstyle 7}$}
\rput(0.75,1.25){$\pmb{\scriptstyle 7}$}
\rput(0.25,1.25){$\pmb{\scriptstyle 9}$}
\rput(0.25,0.75){$\pmb{\scriptstyle 9}$}
\rput(0.25,0.25){$\pmb{\scriptstyle 9}$}
\psline[linewidth=0.5pt,arrowsize=1.5pt 4]{->}(2.75,0.75)(1.5,1)
\rput(3.5,1.2){\bf\small maximal}
\rput(4.1,0.8){\bf\small monochromatic}
\rput(3.3,0.4){\bf\small region}
\psframe[linewidth=2pt,linecolor=black,fillstyle=none,fillcolor=gray,origin={0,0}](2.5,2)(3,3.5)
\rput(2.75,2.25){\color{green}$\pmb{\scriptstyle 7}$}
\rput(2.75,2.75){\color{green}$\pmb{\scriptstyle 7}$}
\rput(2.75,3.25){\color{green}$\pmb{\scriptstyle 7}$}
\rput(2.75,3.75){$\pmb{\scriptstyle 9}$}
\rput(2.25,3.75){$\pmb{\scriptstyle 9}$}
\rput(2.25,3.25){$\pmb{\scriptstyle 9}$}
\rput(2.25,2.75){$\pmb{\scriptstyle 5}$}
\rput(2.75,1.75){$\pmb{\scriptstyle 5}$}
\rput(3.25,1.75){$\pmb{\scriptstyle 5}$}
\rput(3.25,2.25){$\pmb{\scriptstyle 3}$}
\rput(3.25,2.75){$\pmb{\scriptstyle 1}$}
\rput(3.25,3.25){$\pmb{\scriptstyle 3}$}
\rput(3.25,3.75){$\pmb{\scriptstyle 3}$}
\psline[linewidth=0.5pt,arrowsize=1.5pt 4]{->}(2,4.5)(2.75,3.5)
\rput(2,5.5){\bf\small (maximal)}
\rput(2,5.1){\bf\small monochromatic}
\rput(2,4.7){\bf\small rectangle}
\end{pspicture}
\vspace*{-0.1in}
\caption{\label{example}\em An illustration of some communication-complexity definitions.}
\end{wrapfigure}

We have two parties $\party_1$ and $\party_2$, each a binary string, $x_1$ and $x_2$ respectively,
which represents a private value in some set $\cU^{\,\mathrm{in}}$.
The common goal of the two parties is to compute the value $f(x_1,x_2)$ of a given public-knowledge two-variable function $f$.
Before a communication protocol $P$ starts, each $\party_i$ initializes its ``{\em set of maintained inputs}''
$\cU^{\,\mathrm{in}}_i$ to $\cU^{\,\mathrm{in}}$. 
In one step of communication, one party transmits a bit indicating in which of two parts of its input space its private input lies. 
The other party then updates its set of maintained inputs accordingly.
The very last information transmitted in the protocol $P$ contains the value of
$f(x_1,x_2)$. The final transcript of the protocol (\IE, the entire information exchanged) is denoted by $s(x_1,x_2)$.

Denoting the domain of outputs by $\cU^{\,\mathrm{out}}$, any function
$f:\cU^{\,\mathrm{in}}\times\cU^{\,\mathrm{in}}\mapsto\cU^{\,\mathrm{out}}$ can be visualized as
$\left|\cU^{\,\mathrm{in}}\right|\times \left|\cU^{\,\mathrm{in}}\right|$ matrix with entries from $\cU^{\,\mathrm{out}}$
in which the first dimension represents the possible values of $\party_1$, ordered by some permutation $\Pi_1$,
while the second dimension represents the possible values of
$\party_2$, ordered by some permutation $\Pi_2$, and each entry contains the value of
$f$ associated with a particular set of inputs from the two parties.
This matrix will be denoted by $A_{\Pi_1,\Pi_2}(f)$, or sometimes simply by $A$.

We now present the following definitions from~\cite{HN97,FJS10}; see Fig.~\ref{example}
for a geometric illustration.

\begin{Definition}[Regions, partitions]
A region of $A$ is any subset of entries in $A$.
A partition of $A$ is a collection of disjoint regions in $A$
whose union equals to $A$
\end{Definition}

\begin{Definition}[Rectangles, tilings, refinements]
A rectangle in $A$ is a submatrix of $A$.
A tiling of $A$ is a partition of $A$ into rectangles.
A tiling $T_1$ of $A$ is a refinement of another tiling $T_2$ of
$A$ if every rectangle in $T_1$ is contained in some rectangle in $T_2$.
\end{Definition}

\begin{Definition}[Monochromatic, maximal monochromatic and ideal monochromatic partitions]
A region $R$ of $A$ is monochromatic if all entries in $R$ are of the same value.
A monochromatic partition of $A$ is a partition all of whose regions are monochromatic.
A monochromatic region of $A$ is a maximal monochromatic region
if no monochromatic region in $A$ properly contains it. The ideal monochromatic partition of
$A$ is made up of the maximal monochromatic regions.
\end{Definition}

\begin{Definition}[Perfect privacy]
Protocol $P$ achieves perfect privacy if, for every two sets of inputs
$(x_1, x_2)$ and $(x_1', x_2')$ such that $f(x_1, x_2)=f(x_1', x_2')$, it holds that $s(x_1,\ x_2)=s({x'}_1,\ {x'}_2)$.
Equivalently, a protocol $P$ for $f$ achieves perfectly privacy if the monochromatic tiling induced by $P$ is the
ideal monochromatic partition of $A(f)$.
\end{Definition}

\begin{Definition}[Worst case and average case \PAR\ of a protocol $P$]
Let $R^P(x_1,x_2)$ be the monochromatic rectangle containing the cell $A(x_1,x_2)$ induced by $P$,
$R^I(x_1,x_2)$ be the monochromatic region containing the cell $A(x_1,y_1)$ in the ideal monochromatic partition of $A$, and
$\cD$ be a probability distribution over the space of inputs.
Then $P$ has a worst-case \PAR\ of $\aw$
and an average case \PAR\ of $\alpha_{\scriptscriptstyle\cD}$
under distribution $\cD$  provided\footnote{The notation $\displaystyle\Pr_{\scriptscriptstyle\cD}[\mathcal{E}]$ denotes the
probability of an event $\mathcal{E}$ under distribution $\cD$.}
\[
\aw=\!\!\!\!\!\!\!\!\max_{(x_1,x_2)\in\,\cU^{\,\mathrm{in}}\times\cU^{\,\mathrm{in}}}\frac{|\,R^I(x_1,x_2)\,|}{|\,R^P(x_1,x_2)|}
\,\,\,
\mbox{and}
\,\,\,
\alpha_{\scriptscriptstyle\cD}=\!\!\!\!\!\!\!\!\!\!\!\!\sum_{(x_1,x_2)\in\,\cU^{\,\mathrm{in}}\times\cU^{\,\mathrm{in}}}
\!\!\!\!\!\!\!\!\!\!\!\!
\Pr_{\scriptscriptstyle\cD}\left[x_1\,\&\,x_2\right]\frac{\left|R^I(x_1,x_2)\right|}{\left|R^P(x_1,x_2)\right|}
\]
\end{Definition}

\begin{Definition}[\PAR\ for a function]
The worst-case (average-case) \PAR\ for a function $f$ is the minimum, over all protocols $P$ for $f,$ of the worst-case (average-case) \PAR\ of $P$.
\end{Definition}

\paragraph{Extension to Multi-party Computation}
In the multi-party setup, we have $d>2$ parties $\party_1,\party_2,\dots,\party_d$ computing a $d$-ary function
$f:(\cU^{\,\mathrm{in}})^d\mapsto\cU^{\,\mathrm{out}}$. Now, $f$ can be visualized as $\left|\cU^{\,\mathrm{in}}\right|\times\dots\times |\cU^{\,\mathrm{in}}|$ matrix
$A_{\Pi_1,\dots,\Pi_d}(f)$ (or, sometimes simply by $A$) with entries from $\cU^{\,\mathrm{out}}$
in which the $i\tx$ dimension represents the possible values of $\party_i$ ordered by some permutation $\Pi_i$,
and each entry of $A$ contains the value of $f$ associated with a particular set of inputs from the $d$ parties.
Then, all the previous definitions can be naturally adjusted in the obvious manner, \IE,
the input space as a $d$-dimensional space, each party maintains the input partitions of all other $d-1$ parties,
the transcript of the protocol $s$ is a $d$-ary function,
and rectangles are replaced by $d$-dimensional hyper-rectangles (Cartesian product of $d$ intervals).

\subsection{Dissection Protocols and Tiling Functions for Two-party Computation}
\label{setup2}

Often in a communication complexity settings the input of each party has a natural ordering, \EG, the set of input of a party from $\big\{0,1\big\}^k$ can represent the numbers
$0,1,2,\dots,2^k-1$ (as is the case when computing the maximum/minimum of two inputs, in the millionaires problem, in second-price auctions, and more).
When designing protocols for such environments, a natural restriction is to only the allow protocols such that each party asks questions of the
form ``{\em Is your input between $a$ and $b$ (in this natural order over possible inputs)}?'',
where $a,b\in\big\{0,1\big\}^k$.
Notice that after applying an appropriate permutation to the inputs, such a protocol divides the input space into two (not necessarily equal) halves.
Below, we formalize these types of protocols as ``{\em dissection protocols}''.

\begin{Definition}[contiguous subset of inputs]
Given a permutation $\Pi$ of $\{0,1\}^k$, let
$\prec_{\scriptscriptstyle\,\Pi}$ denote the total order over $\{0,1\}^k$ that $\Pi$ induces, \IE,
$\forall\, a,b\in \{0,1\}^k$, $a \prec_{\scriptscriptstyle\Pi} b$ provided $b$ comes after $a$ in $\Pi$.
Then, $I\subseteq \{0,1\}^k$ {\em contiguous} with respect to $\Pi$ if
$
\,\,\forall\,a,b\!\in\! I,\,\forall\,c\!\in\! \big\{0,1\big\}^k\colon\, a\prec_{\scriptscriptstyle\Pi} c\prec_{\scriptscriptstyle\Pi} b\Longrightarrow c\in I
$.
\end{Definition}

\begin{Definition}[dissection protocol]
Given a function $f:\{0,1\}^k\times\{0,1\}^k\mapsto \{0,1\}^t$ and permutations $\Pi_1,\Pi_2$ of $\{0,1\}^k$,
a protocol for $f$ is a {\em dissection protocol} with respect to
$(\Pi_1,\Pi_2)$ if, at each communication step, the maintained subset of inputs of each $\party_i$ is contiguous with respect to $\Pi_i$.
\end{Definition}

Observe that {\em every} protocol $P$ can be regarded as a dissection protocol with respect to {\em some} permutations over inputs by simply
constructing the permutation so that it is consistent with the way $P$ updates the maintained sets of inputs. However, {\em not} every protocol is a
dissection protocol with respect to {\em specific} permutations. Consider, for example, the case that both $\Pi_1$ and $\Pi_2$ are the
permutation over $\{0,1\}^k$ that orders the elements from lowest to highest binary values.
Observe that a protocol that is a dissection protocol with respect to these permutations {\em cannot} ask questions of the
form ``Is your input odd or even?'', for these questions partition the space of inputs into {\em non-contiguous} subsets with respect to $(\Pi_1,\Pi_2)$.

A special case of interest of the dissection protocol is the ``bisection type'' protocols that have been investigated
in the literature in many contexts~\cite{FJS10,muller2}.

\begin{Definition}[bisection, $c$-bisection and bounded-bisection protocols]\label{bidef}
For a constant $c\in \left.\left[\frac{1}{2},1\right.\right)$, a dissection protocol with respect to the permutations $(\Pi_1,\Pi_2)$ is
called a $c$-bisection protocol provided at each communication step
each $\party_i$ partitions its input space of size $z$ into two halves of size
$c\,z$ and $(1-c)\,z$. A bisection protocol is simply a $\frac{1}{2}$-bisection protocol.
For an integer valued function $g(k)$ such that $0\leq g(k)\leq k$,
$\mbox{bounded-bisection}_{g(k)}$ is the protocol that runs a bisection protocol with $g(k)$ bisection operations
followed by a protocol (if necessary) in which each $\party_i$ repeatedly partitions its input space into
two halves one of which is of size exactly one.
\end{Definition}

We next introduce the concept of {\em tiling} functions.

\begin{Definition}[tiling and non-tiling functions]\label{tiledef}
A function $f:\{0,1\}^k\times \{0,1\}^k\mapsto \{0,1\}^t$ is
called a tiling function with respect to two permutations $(\Pi_1,\Pi_2)$ of
$\{0,1\}^k$ if the monochromatic regions in $A_{\Pi_1,\Pi_2}(f)$ form a tiling, and
the number of monochromatic regions in this tiling is denote by $\rr_f\left(\Pi_1,\Pi_2\right)$.
Conversely, $f$ is a non-tiling function if
$f$ is not a tiling function with respect to every pair of permutations $(\Pi_1,\Pi_2)$ of $\{0,1\}^k$.
\end{Definition}

\begin{figure}[htbp]
\begin{minipage}[b]{6.5cm}
\begin{pspicture}(-0.3,0.2)(6,4)
\hspace*{0.1in}
\psset{xunit=0.5cm,yunit=0.5cm}
\rput(1.5,0.6){$\scriptstyle\pmb{00}$}
\rput(2.5,0.6){$\scriptstyle\pmb{01}$}
\rput(3.5,0.6){$\scriptstyle\pmb{10}$}
\rput(4.5,0.6){$\scriptstyle\pmb{11}$}
\rput(0.6,0.6){$\pmb{\Pi_1}$}
\rput(3,0){\bf (a)}
\rput(0.5,1.5){$\scriptstyle\pmb{00}$}
\rput(0.5,2.5){$\scriptstyle\pmb{01}$}
\rput(0.5,3.5){$\scriptstyle\pmb{10}$}
\rput(0.5,4.5){$\scriptstyle\pmb{11}$}
\rput(-0.4,3.1){$\pmb{\Pi_2}$}
\rput(2.5,3.5){$\pmb{1}$}
\rput(2.5,2.5){$\pmb{1}$}
\rput(2.5,1.5){$\pmb{2}$}
\rput(3.5,1.5){$\pmb{2}$}
\rput(4.5,1.5){$\pmb{3}$}
\rput(4.5,2.5){$\pmb{3}$}
\rput(4.5,3.5){$\pmb{4}$}
\rput(3.5,3.5){$\pmb{4}$}
\rput(3.5,2.5){$\pmb{5}$}
\rput(1.5,4.5){$\pmb{6}$}
\rput(1.5,3.5){$\pmb{7}$}
\rput(1.5,2.5){$\pmb{8}$}
\rput(1.5,1.5){$\pmb{9}$}
\rput(2.5,4.5){$\pmb{{10}}$}
\rput(3.5,4.5){$\pmb{{11}}$}
\rput(4.5,4.5){$\pmb{{12}}$}
\psframe[linewidth=2pt,linecolor=gray,origin={0,0}](1,4)(2,5)
\psframe[linewidth=2pt,linecolor=gray,origin={0,-1}](1,4)(2,5)
\psframe[linewidth=2pt,linecolor=gray,origin={0,-2}](1,4)(2,5)
\psframe[linewidth=2pt,linecolor=gray,origin={0,-3}](1,4)(2,5)
\psframe[linewidth=2pt,linecolor=gray,origin={1,0}](1,4)(2,5)
\psframe[linewidth=2pt,linecolor=gray,origin={2,0}](1,4)(2,5)
\psframe[linewidth=2pt,linecolor=gray,origin={3,0}](1,4)(2,5)
\psframe[linewidth=3pt,origin={0,0}](2,2)(3,4)
\psframe[linewidth=3pt,origin={2,-1}](2,2)(3,4)
\psframe[linewidth=3pt,origin={0,0}](2,1)(4,2)
\psframe[linewidth=3pt,origin={1,2}](2,1)(4,2)
\psframe[linewidth=3pt,origin={0,0}](3,2)(4,3)
\hspace*{-0.1in}
\hspace*{3.6cm}
\rput(1.5,0.6){$\scriptstyle\pmb{01}$}
\rput(2.5,0.6){$\scriptstyle\pmb{00}$}
\rput(3.5,0.6){$\scriptstyle\pmb{10}$}
\rput(4.5,0.6){$\scriptstyle\pmb{11}$}
\rput(0.5,0.6){$\pmb{\Pi_1'}$}
\rput(3,0){\bf (b)}
\rput(0.5,1.5){$\scriptstyle\pmb{00}$}
\rput(0.5,2.5){$\scriptstyle\pmb{01}$}
\rput(0.5,3.5){$\scriptstyle\pmb{10}$}
\rput(0.5,4.5){$\scriptstyle\pmb{11}$}
\rput(-0.4,3.1){$\pmb{\Pi_2'}$}
\rput(1.5,3.5){$\pmb{1}$}
\rput(1.5,2.5){$\pmb{1}$}
\rput(1.5,1.5){$\pmb{2}$}
\rput(3.5,1.5){$\pmb{2}$}
\rput(4.5,1.5){$\pmb{3}$}
\rput(4.5,2.5){$\pmb{3}$}
\rput(4.5,3.5){$\pmb{4}$}
\rput(3.5,3.5){$\pmb{4}$}
\rput(3.5,2.5){$\pmb{5}$}
\rput(1.5,4.5){$\pmb{{10}}$}
\rput(3.5,4.5){$\pmb{{11}}$}
\rput(4.5,4.5){$\pmb{{12}}$}
\rput(2.5,4.5){$\pmb{6}$}
\rput(2.5,3.5){$\pmb{7}$}
\rput(2.5,2.5){$\pmb{8}$}
\rput(2.5,1.5){$\pmb{9}$}
\psframe[linewidth=2pt,linecolor=gray,origin={0,0}](1,4)(2,5)
\psframe[linewidth=2pt,linecolor=gray,origin={1,0}](1,4)(2,5)
\psframe[linewidth=2pt,linecolor=gray,origin={2,0}](1,4)(2,5)
\psframe[linewidth=2pt,linecolor=gray,origin={3,0}](1,4)(2,5)
\psframe[linewidth=2pt,linecolor=gray,origin={1,0}](1,4)(2,5)
\psframe[linewidth=2pt,linecolor=gray,origin={1,-1}](1,4)(2,5)
\psframe[linewidth=2pt,linecolor=gray,origin={1,-2}](1,4)(2,5)
\psframe[linewidth=2pt,linecolor=gray,origin={1,-3}](1,4)(2,5)
\psframe[linewidth=3pt,origin={-1,0}](2,2)(3,4)
\psframe[linewidth=3pt,origin={1,-1}](2,2)(3,3)
\psframe[linewidth=3pt,origin={1,-1}](2,3)(3,4)
\psframe[linewidth=3pt,origin={1,-1}](2,2)(3,4)
\psframe[linewidth=3pt,origin={-1,0}](2,1)(3,2)
\psframe[linewidth=3pt,origin={1,2}](2,1)(4,2)
\psframe[linewidth=3pt,origin={1,0}](3,2)(4,3)
\psframe[linewidth=3pt,origin={1,-1}](3,2)(4,3)
\end{pspicture}
\caption{\label{perm}\em A tiling function with respect to different permutation pairs
$(\Pi_1,\Pi_2)$ and $(\Pi_1',\Pi_2')$ inducing different numbers of
monochromatic rectangles.}
\end{minipage}
\hspace*{1cm}
\begin{minipage}[b]{5.5cm}
\begin{pspicture}(-0.4,0)(5,3)
\psset{xunit=1.3cm,yunit=0.65cm}
\psframe[linewidth=0pt,linecolor=gray,fillstyle=solid,fillcolor=white,origin={0,0}](0,0)(1.5,3)
\psframe[linewidth=0pt,linecolor=gray,fillstyle=solid,fillcolor=lightgray,origin={0,0}](0,2)(0.5,3)
\psframe[linewidth=0pt,linecolor=gray,fillstyle=solid,fillcolor=gray,origin={0,0}](0.5,2)(1,3)
\psframe[linewidth=0pt,linecolor=gray,fillstyle=solid,fillcolor=lightgray,origin={0,0}](0.5,1)(1,2)
\psframe[linewidth=0pt,linecolor=gray,fillstyle=solid,fillcolor=gray,origin={0,0}](0.5,0)(1,1)
\psframe[linewidth=0pt,linecolor=gray,fillstyle=solid,fillcolor=lightgray,origin={0,0}](0,0)(0.5,1)
\psframe[linewidth=0pt,linecolor=gray,fillstyle=solid,fillcolor=black,origin={0,0}](1,0)(1.5,3)
\psframe[linewidth=0pt,linecolor=gray,fillstyle=solid,fillcolor=black,origin={0,0}](0,1)(0.5,2)
\rput(0.4,-0.4){$\pmb{\scriptstyle\Pi_1}$}
\rput(-0.3,1){$\pmb{\scriptstyle\Pi_2}$}
\hspace*{3cm}
\psframe[linewidth=0pt,linecolor=gray,fillstyle=solid,fillcolor=white,origin={0,0}](0,0)(1.5,3)
\psframe[linewidth=0pt,linecolor=gray,fillstyle=solid,fillcolor=gray,origin={0,0}](0,2)(0.5,3)
\psframe[linewidth=0pt,linecolor=gray,fillstyle=solid,fillcolor=lightgray,origin={0,0}](0.5,2)(1,3)
\psframe[linewidth=0pt,linecolor=gray,fillstyle=solid,fillcolor=lightgray,origin={0,0}](0,1)(0.5,2)
\psframe[linewidth=0pt,linecolor=gray,fillstyle=solid,fillcolor=lightgray,origin={0,0}](0.5,0)(1,1)
\psframe[linewidth=0pt,linecolor=gray,fillstyle=solid,fillcolor=gray,origin={0,0}](0,0)(0.5,1)
\psframe[linewidth=0pt,linecolor=gray,fillstyle=solid,fillcolor=black,origin={0,0}](1,0)(1.5,3)
\psframe[linewidth=0pt,linecolor=gray,fillstyle=solid,fillcolor=black,origin={0,0}](0.5,1)(1,2)
\rput(0.4,-0.4){$\pmb{\scriptstyle\Pi_1'}$}
\rput(-0.3,1){$\pmb{\scriptstyle\Pi_2'}$}
\end{pspicture}
\caption{\label{tot}\em Tilability depends on $\Pi_1$ and $\Pi_2$.}
\end{minipage}
\end{figure}


For example,
$f(x_1,\dots,x_k,y_1,\dots,y_k)\equiv\sum_{i=1}^k\left(x_i+y_i\right)\pmod{2}$
is a tiling function with respect to $(\Pi_1,\Pi_2)$ with $\rr_f\left(\Pi_2,\Pi_2\right)=4$,
where each $\Pi_i$ orders its inputs $(z_1,\dots,z_k)$ in increasing order of $\sum_{i=1}^kz_i\pmod{2}$.
Note that a function $f$ that is tiling function with respect to permutations $(\Pi_1,\Pi_2)$
may not be a tiling function
with respect to a different set of permutations $(\Pi_1',\Pi_2')$; see Fig.~\ref{tot}.
Also, a function $f$ can be a tiling function with respect to two distinct permutation pairs $(\Pi_1,\Pi_2)$ and
$(\Pi_1',\Pi_2')$, and the number of monochromatic regions in the two cases
differ; see Fig.~\ref{perm}.
Thus, indeed we need $\Pi_1$ and $\Pi_2$ in the definition of tiling functions and $\rr_f$.


\paragraph{Extensions to Multi-party Computation}
For the multi-party computation model involving $d>2$ parties,
the $d$-ary tiling function $f$ has a permutation $\Pi_i$ of $\{0,1\}^k$ for each $i\tx$ argument of $f$ (or, equivalently
for each $\party_i$). A dissection protocol is generalized to a ``round robin'' dissection protocol in the following manner.
In one ``mega'' round of communications, parties communicate in a fixed order, say $\party_1,\party_2,\dots,\party_d$,
and the mega round is repeated if necessary. Any communication by any party is made available to
{\em all} the other parties.  Thus, each communication of the dissection protocol
partitions a $d$-dimensional space by an appropriate {\em set} of $(d-1)$-dimensional hyperplanes,
where the missing dimension in the hyperplane correspond to the index of the party communicating.

\section{Two-party Dissection Protocol for Tiling Functions}

\subsection{Boolean Tiling Functions}
\label{bool}

\begin{Lemma}\label{thm:boolean}
Any Boolean tiling function $f\colon \{0,1\}^k\times \{0,1\}^k\mapsto \{0,1\}$ with respect to some two permutations $(\Pi_1,\Pi_2)$
can be computed in a perfectly privacy-preserving manner by a dissection protocol with respect to the same permutations $(\Pi_1,\Pi_2)$.
\end{Lemma}

\begin{wrapfigure}[12]{R}{1.7in}
\begin{pspicture}(-0.7,0)(4,3.5)
\psset{xunit=0.8cm,yunit=0.8cm}
\psframe[linewidth=1pt,linecolor=black](0,0)(4,4)
\psline[linewidth=1pt,linecolor=black,linestyle=dotted](4,0)(4.5,0)(4.5,4)(4,4)
\rput(-0.2,0.2){$\pmb{1}$}
\rput(-0.2,1.25){$\pmb{i}$}
\rput(-0.5,1.75){$\pmb{i+1}$}
\rput(-0.2,3.8){$\pmb{m}$}
\rput(0.1,4.25){$\pmb{1}$}
\rput(3.7,4.2){$\pmb{q}$}
\rput(4.3,4.2){$\scriptstyle \pmb{q+1}$}
\psline[linewidth=1pt,linecolor=black](0,1.5)(4,1.5)
\psline[linewidth=1pt,linecolor=black](0,1)(4,1)
\psline[linewidth=1pt,linecolor=black](0,2)(4,2)
\psline[linewidth=1pt,linecolor=black](3.5,4)(3.5,0)
\psframe[linewidth=0pt,linecolor=black,fillstyle=solid,fillcolor=black](4,1)(4.5,2)
\psframe[linewidth=0pt,linecolor=black,fillstyle=solid,fillcolor=black](3.5,1)(4,1.5)
\psframe[linewidth=0pt,linecolor=black,fillstyle=solid,fillcolor=lightgray](3.5,1.5)(4,2)
\end{pspicture}
\vspace*{-0.3in}
\caption{\em\label{some1}This configuration cannot happen in Case~{\bf 2}.}
\end{wrapfigure}
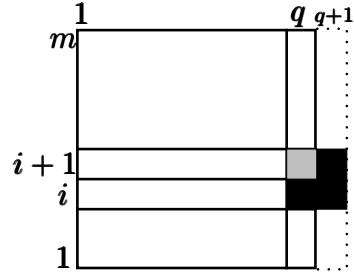

\noindent
{\em Proof}.\
For any $m\times n$ Boolean matrix $A$ with rows and columns indexed by $1,2,\dots,m$ and $1,2,\dots,n$, respectively,
let the notation $A[i_1,i_2,j_1,j_2]$ denote the submatrix of $A$ consisting of rows $i_1,i_1+1,\dots,i_2$ and
columns $j_1,j_1+1,\dots,j_2$.
Assume $m,n\geq 2$ and suppose that the zeroes and ones in the matrix $A$ form a tiling. We claim
that there must exist an index $i\in\{1,2,\dots,m-1\}$ such that the partition of $A$ into the two submatrices
$A[1,i,1,n]$ and $A[i+1,m,1,n]$ does not split any tile,
or that there must exist an index $j\in\{1,2,\dots,n-1\}$ such that the partition of $A$ into the two sub-matrices
$A[1,m,1,j]$ and $A[1,m,j+1,n]$ does not split any tile.
This claim, applied recursively on each submatrix of $A$, will prove Lemma~\ref{thm:boolean}.

We prove our claim by induction on $n$. The basis case of $n=2$ follows trivially.
Suppose that our claim is true for all $n\in \{2,\dots,q\}$ and consider the case of $n=q+1$.

\vspace*{0.1in}
\noindent
{\bf Case 1:}
there exists an index $j\in\{1,2,\dots,q-1\}$ such that the partition of $A[1,m,1,q]$ into the two sub-matrices
$A[1,m,1,j]$ and $A[1,m,j+1,q]$ does not split any tile. Then, the same index $j$ works for $A[1,m,1,q+1]$ also.

\vspace*{0.1in}
\noindent
{\bf Case 2:}
there is no such index $j$ as in Case~{\bf 1} above, but
there exists an index $i\in\{1,2,\dots,m-1\}$ such that the partition of $A[1,m,1,q]$ into the two submatrices
$A[1,i,1,q]$ and $A[i+1,m,1,q]$ does not split any tile.
Suppose that the index $i$ does split a tile in the partition
$A[1,i,1,q+1]$ and $A[i+1,m,1,q+1]$ of $A[1,m,1,q+1]$. Then, we must have the situation as shown in Fig.~\ref{some1},
which shows that the zeroes and ones of $A[1,m,1,q+1]$ do not form a tiling.
\hfill{\Pisymbol{pzd}{113}}

\begin{remark}
As Fig.~\ref{some1} shows, the claim of Lemma~\ref{thm:boolean} is false if $f$ outputs three values.
\end{remark}

\subsection{Average and Worst Case \PAR\ for Non-Boolean Tiling Functions}
\label{par-theorems}

\begin{figure}[ht]
\begin{minipage}[b]{4.34cm}
\begin{pspicture}(-3.7,-4)(0,2.7)
\psset{xunit=1.2cm,yunit=1.2cm}
\psframe[linewidth=0pt,linecolor=gray,fillstyle=solid,fillcolor=black](0,0)(2,0.3)
\psframe[linewidth=0pt,linecolor=gray,fillstyle=solid,fillcolor=gray](0,0.3)(2,2)
\psline[linewidth=0.5pt,arrowsize=1.5pt 2]{|<->|}(-0.2,0)(-0.2,0.3)
\rput(-0.4,0.15){$\scriptstyle 1$}
\psline[linewidth=0.5pt,arrowsize=1.5pt 4]{|<->|}(-0.2,0.3)(-0.2,2)
\rput(-0.6,1.15){$\scriptstyle 2^k-1$}
\psline[linewidth=0.5pt,arrowsize=1.5pt 4]{|<->|}(0,2.2)(2,2.2)
\rput(1,2.4){$\scriptstyle 2^k$}
\rput(2.4,1){\Large $\pmb{f_1}$}
\psset{origin={0,-2.8}}
\psframe[linewidth=0pt,linecolor=gray,fillstyle=solid,fillcolor=gray](0,0)(2,2)
\psframe[linewidth=0pt,linecolor=gray,fillstyle=solid,fillcolor=black](0,1.3)(2,1.6)
\psline[linewidth=0.5pt,arrowsize=1.5pt 2]{|<->|}(-0.2,1.3)(-0.2,1.6)
\rput(-0.4,-1.35){$\scriptstyle 1$}
\psline[linewidth=0.5pt,arrowsize=1.5pt 2]{|<->|}(-0.2,1.6)(-0.2,2)
\rput(-0.9,-1){$\scriptstyle 2^{k-g(k)-1}$}
\psline[linewidth=0.5pt,arrowsize=1.5pt 4]{|<->|}(-0.2,0)(-0.2,1.3)
\rput(-0.9,-2.15){$\scriptstyle 2^k-2^{k-g(k)}$}
\psline[linewidth=0.5pt,arrowsize=1.5pt 4]{|<->|}(0,2.2)(2,2.2)
\rput(1,-0.4){$\scriptstyle 2^k$}
\rput(2.4,-1.8){\Large $\pmb{f_2}$}
\end{pspicture}
\end{minipage}
\\
\begin{minipage}[b]{5in}
\begin{tabular}{c|c|c}\toprule
                                            &  number of                      & average case      \\
 protocol                                    & steps            of            & \PAR\ for         \\
                                             & communication                  &      distribution $\UD$ \\ \hline
$c$-bisection on $f_1$                      & $k/\log_2\frac{1}{c}$          & $k$ \\
                        $(\,c\in[1/2,1)\,)$  &                                 &      \\ \hline
$\mbox{bounded-bisection}_{g(k)}$ on $f_2$  & $g(k)+2^{k-g(k)}-1$            & $g(k)+2^{k-g(k)-1}-1$ \\
$(\,1\leq g(k)\leq k\,)$                    & \\
\bottomrule
\multicolumn{3}{c}{} \\
\multicolumn{3}{c}{} \\
\multicolumn{3}{c}{} \\
\multicolumn{3}{c}{} \\
\end{tabular}
\end{minipage}
\vspace*{-0.8in}
\caption{\em\label{badcase} Functions $f_1$ and $f_2$ with $\rr_{f_1}(\Pi_1,\Pi_2)=\rr_{f_2}(\Pi_1,\Pi_2)=2$.
The bisection-type protocols fail to achieve a good average-case \PAR\ on them.}
\end{figure}

Let $f:\{0,1\}^k\times \{0,1\}^k\mapsto \{0,1\}^t$ be a given tiling function with respect to permutations $(\Pi_1,\Pi_2)$.
Neither the $c$-bisection nor the bounded-bisection protocol performs well in terms of average \PAR\
on arbitrary tiling functions; see Fig.~\ref{badcase} for an illustration.
In this section, we show that {\em any} tiling function $f$ admits a dissection protocol
that has a {\em small constant} average case \PAR.
Moreover, we show that this result {\em cannot} be extended to the case of
worst-case {\sc Par}s.

\subsubsection{Constant Average-case \PAR\ for Non-Boolean Functions}

Let $\UD$ denote the uniform distribution over all input pairs.
We define the notion of a $c$-approximate uniform distribution $\UD^{\thicksim\, c}$; note that
$\UD^{\thicksim\, 0}\,\equiv\,\UD$.

\begin{Definition}[$c$-approximate uniform distribution]
A $c$-approximate uniform distribution $\UD^{\thicksim\, c}$ is a distribution in which
the probabilities of the input pairs are close to that for the uniform distribution as a linear function of $c$, namely
\[
\max_{(\mathbf{x},\mathbf{y}),\,(\mathbf{x'},\mathbf{y'})\in\{0,1\}^k\times \{0,1\}^k}
\left\lvert\,
{\Pr_{\,\scriptscriptstyle \UD^{\thicksim\,c}}\left[\mathbf{x}\,\&\,\mathbf{y}\right]-\Pr_{\,\scriptscriptstyle \UD^{\thicksim\,c}}\left[\mathbf{x'}\,\&\,\mathbf{y'}\right] }
\,\right\rvert
\leq c\,2^{-2k}
\]
\end{Definition}

\begin{Theorem}\label{x}~\\
\noindent
{\bf (a)}
A tiling function $f$ with respect to permutations $(\Pi_1,\Pi_2)$ admits a dissection protocol $P$ with respect
to the same permutations $(\Pi_1,\Pi_2)$ using at most $4\,\rr_f(\Pi_1,\Pi_2)$ communication steps such that
$\alpha_{\scriptscriptstyle\UD^{\thicksim\,c}}\leq 4+4\,c$.

\vspace*{0.1in}
\noindent
{\bf (b)}
For all $0\leq c<1$, there exists a tiling function $f\colon\{0,1\}^2\times\{0,1\}^2\mapsto\{0,1\}^4$ such that,
for any two permutations $(\Pi_1,\Pi_2)$ of $\{0,1\}^2$,
every dissection protocol with respect to $(\Pi_1,\Pi_2)$ using any number of communication steps has
$\alpha_{\scriptscriptstyle\UD^{\thicksim\,c}}\geq \frac{9+c}{8}$.
\end{Theorem}

\noindent
{\em Proof}.\
Let $\cS=\{S_1,S_2,\dots,S_{\rr_f}\}$ be the set of $\rr_f=\rr_f(\Pi_1,\Pi_2)$ ideal monochromatic rectangles in
the tiling of $f$ induced by the permutations $(\Pi_1,\Pi_2)$ and
consider a protocol $P$ that is a dissection
protocol with respect to $(\Pi_1,\Pi_2)$. Suppose that the ideal monochromatic rectangle $S_i\in\cS$
has $y_i$ elements, and $P$ partitions this rectangle
into $t_i$ rectangles $S_{i,1},\dots,S_{i,t_i}$ having $z_{i,1},\dots,z_{i,t_i}$ elements, respectively.
Then, using the definition of $\alpha_{\scriptscriptstyle\UD}$ it follows that
\begin{multline*}
\alpha_{\scriptscriptstyle\UD}
=
\sum_{(x_1,x_2)\in\cU\times\cU}\Pr_{\UD}\left[x_1\,\&\,x_2\right]\frac{\left|R^I(x_1,x_2)\right|}{\left|R^P(x_1,x_2)\right|}
\\
=
\sum_{i=1}^{\rr_f} \sum_{j=1}^{t_i} \!\!\!\!\!\!\!\!\!\!\!\!\!\! \sum_{\,\,\,\,\,\,\,\,\,\,\,\,\,\,\,\,\,\,(x_1,x_2)\in S_{i,j}}\!\!\!\!\!\!\!\!\!\!\!\!\!\!\!
\Pr_{\UD}[x_1\,\&\,x_2]\,\frac{y_i}{z_{i,j}}
=
\sum_{i=1}^{\rr_f} \sum_{j=1}^{t_i}
\frac{y_i}{2^{2k}}
=
\sum_{i=1}^{\rr_f}
\frac{t_i\,y_i}{2^{2k}}
\end{multline*}
Similarly, it follows that
\begin{gather*}
\alpha_{\scriptscriptstyle\UD^{\thicksim\,c}}
\leq
\sum_{i=1}^{\rr_f} \sum_{j=1}^{t_i} \!\!\!\!\!\!\!\!\!\!\!\!\!\! \sum_{\,\,\,\,\,\,\,\,\,\,\,\,\,\,\,\,\,\,(x_1,x_2)\in S_{i,j}}\!\!\!\!\!\!\!\!\!\!\!\!\!\!\!
\frac{1+c}{2^{2k}}\times\frac{y_i}{z_{i,j}}
=
\sum_{i=1}^{\rr_f} \sum_{j=1}^{t_i}
\frac{(1+c)\,y_i}{2^{2k}}
=
\sum_{i=1}^{\rr_f}
\frac{(1+c)\,t_i\,y_i}{2^{2k}}
\end{gather*}

\begin{figure}[htbp]
\begin{minipage}[b]{2.6in}
\begin{pspicture}(-0.5,-2.7)(4.6,5)
\psset{xunit=1cm,yunit=1cm}
\rput(2.5,3.5){$\pmb{\frac{1+c}{16}}$}
\rput(2.5,2.5){$\pmb{\frac{1+c}{16}}$}
\rput(2.5,1.5){$\pmb{\frac{1+c}{16}}$}
\rput(3.5,1.5){$\pmb{\frac{1+c}{16}}$}
\rput(4.5,1.5){$\pmb{\frac{1+c}{16}}$}
\rput(4.5,2.5){$\pmb{\frac{1+c}{16}}$}
\rput(4.5,3.5){$\pmb{\frac{1+c}{16}}$}
\rput(3.5,3.5){$\pmb{\frac{1+c}{16}}$}
\rput(3.5,2.5){$\pmb{\frac{1-c}{16}}$}
\rput(1.5,4.5){$\pmb{\frac{1-c}{16}}$}
\rput(1.5,3.5){$\pmb{\frac{1-c}{16}}$}
\rput(1.5,2.5){$\pmb{\frac{1-c}{16}}$}
\rput(1.5,1.5){$\pmb{\frac{1-c}{16}}$}
\rput(2.5,4.5){$\pmb{{\frac{1-c}{16}}}$}
\rput(3.5,4.5){$\pmb{{\frac{1-c}{16}}}$}
\rput(4.5,4.5){$\pmb{{\frac{1-c}{16}}}$}
\psframe[linewidth=2pt,linecolor=black,origin={0,0}](1,4)(2,5)
\psframe[linewidth=2pt,linecolor=black,origin={0,-1}](1,4)(2,5)
\psframe[linewidth=2pt,linecolor=black,origin={0,-2}](1,4)(2,5)
\psframe[linewidth=2pt,linecolor=black,origin={0,-3}](1,4)(2,5)
\psframe[linewidth=2pt,linecolor=black,origin={1,0}](1,4)(2,5)
\psframe[linewidth=2pt,linecolor=black,origin={2,0}](1,4)(2,5)
\psframe[linewidth=2pt,linecolor=black,origin={3,0}](1,4)(2,5)
\psframe[linewidth=2pt,origin={0,0}](2,2)(3,4)
\psframe[linewidth=2pt,origin={2,-1}](2,2)(3,4)
\psframe[linewidth=2pt,origin={0,0}](2,1)(4,2)
\psframe[linewidth=2pt,origin={1,2}](2,1)(4,2)
\psframe[linewidth=2pt,origin={0,0}](3,2)(4,3)
\end{pspicture}

\comment{
\begin{pspicture}(-2.5,-2)(2.6,3)
\hspace*{-0.4in}
\psset{xunit=0.6cm,yunit=0.6cm}
\pspolygon[linewidth=1pt,linecolor=gray,fillstyle=crosshatch,hatchcolor=gray](1,5)(1,5.3)(4.3,5.3)(4.3,2)(4,2)(4,5)
\psline[linewidth=2pt](1,2)(1,5)
\psline[linewidth=2pt](1,5)(4,5)
\psline[linewidth=2pt](4,5)(4,2)
\psline[linewidth=2pt](4,2)(1,2)
\psline[linewidth=2pt](2,2)(2,4)
\psline[linewidth=2pt](1,4)(3,4)
\psline[linewidth=2pt](3,5)(3,3)
\psline[linewidth=2pt](4,3)(2,3)
\psline[linewidth=1pt,arrowsize=1.5pt 4]{|<->|}(0.6,2)(0.6,4)
\rput(-0.8,3){{$\scriptstyle\pmb{\left\lfloor\left(\frac{2}{3}\right)2^k\right\rfloor}$}}
\psline[linewidth=1pt,arrowsize=1pt 3]{|<->|}(0.6,4)(0.6,5)
\rput(-0.9,4.5){$\scriptstyle\pmb{\left\lfloor\left(\frac{1}{3}\right)2^k\right\rfloor}$}
\psline[linewidth=1pt,arrowsize=1.5pt 4]{|<->|}(4.6,2)(4.6,5.3)
\rput(5.1,3.7){{$\scriptstyle\pmb{2^k}$}}
\psline[linewidth=1pt,arrowsize=1.5pt 4]{|<->|}(1,5.6)(4.3,5.6)
\rput(2.6,6){{$\scriptstyle\pmb{2^k}$}}
\end{pspicture}
}
\vspace*{-1.3in}
\caption{\label{notile}\em Example for $\alpha_{\scriptscriptstyle\UD^{\thicksim\,c}}\geq \frac{9+c}{8}$. The tiles are
shown by thick black lines. The numbers shown at each cell is the associated probability of that input.}
\end{minipage}
\hspace*{0.2in}
\begin{minipage}[b]{2.4in}
\begin{pspicture}(-0.7,-1.4)(5.4,3.4)
\psset{xunit=0.5cm,yunit=0.5cm}
\psframe[linewidth=1pt,linecolor=black](0,0)(4,4)
\psframe[linewidth=1pt,linecolor=black](0,0)(1,1)
\psframe[linewidth=1pt,linecolor=black](1,0)(2,3)
\psframe[linewidth=1pt,linecolor=black](0,3)(3,4)
\psframe[linewidth=1pt,linecolor=black](3,4)(3,0)
\psline[linewidth=4pt,linecolor=gray,linestyle=dotted](1,-1)(1,5)
\rput(1,5.3){\em a}
\rput(1,-1.3){\em a}
\psline[linewidth=4pt,linecolor=gray,linestyle=dotted](1,3)(-1,3)
\rput(-1.3,3){\em b}
\psline[linewidth=4pt,linecolor=gray,linestyle=dotted](1,1)(-1,1)
\rput(-1.3,1){\em c}
\psline[linewidth=4pt,linecolor=gray,linestyle=dotted](1,3)(5,3)
\rput(5.3,3){\em d}
\psline[linewidth=4pt,linecolor=gray,linestyle=dotted](3,3)(3,5)
\rput(3,5.3){\em e}
\psline[linewidth=4pt,linecolor=gray,linestyle=dotted](3,3)(3,-1)
\rput(3,-1.3){\em f}
\psline[linewidth=4pt,linecolor=gray,linestyle=dotted](2,3)(2,-1)
\rput(2,-1.3){\em g}
%
\pscircle(8,5){0.2}
\rput(8,5){\em a}
\pscircle(7,3){0.2}
\rput(7,3){\em b}
\pscircle(9,3){0.2}
\rput(9,3){\em d}
\pscircle(6,1){0.2}
\rput(6,1){\em c}
\pscircle(8,1){0.2}
\rput(8,1){\em e}
\pscircle(10,1){0.2}
\rput(10,1){\em f}
\pscircle(9,-1){0.2}
\rput(9,-1){\em g}
%
\psline[linewidth=0.5pt,linecolor=black](7.8,4.7)(7.2,3.3)
\psline[linewidth=0.5pt,linecolor=black](8.2,4.7)(8.8,3.3)
\psline[linewidth=0.5pt,linecolor=black](6.8,2.7)(6.2,1.3)
\psline[linewidth=0.5pt,linecolor=black](8.8,2.7)(8.2,1.3)
\psline[linewidth=0.5pt,linecolor=black](9.2,2.7)(9.8,1.3)
\psline[linewidth=0.5pt,linecolor=black](9.8,0.7)(9.2,-0.7)
\end{pspicture}
\vspace*{-0.5in}
\caption{\label{bsp-fig}\BSP\ and \BSP-tree.}
\end{minipage}
\end{figure}

A binary space partition (\BSP) for a collection of {\em disjoint} rectangles in the two-dimensional plane
is defined as follows. The plane is divided into two parts by cutting rectangles with a line if necessary.
The two resulting parts of the plane are divided recursively in a similar manner;
the process continues until at most one fragment of the original rectangles remains in any part of the plane.
This division process can be naturally represented as a binary tree (\BSP-tree) where a node represents
a part of the plane and stores the cut that splits the plane into two parts that its two children represent and each leaf of the \BSP-tree
represents the final partitioning of the plane by storing at most one fragment of an input rectangle; see Fig.~\ref{bsp-fig}
for an illustration.
The {\em size} of a \BSP\ is the {\em number of leaves} in the \BSP-tree.
The following result is known.

\begin{fact}{\rm\cite{dAF92}}\label{bsp}
Assume that we have a set $\cS$ of disjoint axis-parallel rectangles in the plane. Then,
there is a \BSP\ of $\cS$ such that
every rectangle in $\cS$ is partitioned into at most $4$ rectangles.\footnote{The stronger bounds by Berman, DasGupta and Muthukrishnan~\cite{BDM02}
apply to {\em average} number of fragments only.}
\end{fact}

\begin{figure}[t]
\hspace*{0.5in}
\begin{pspicture}(1,1)(16,8)
\psset{xunit=0.8cm,yunit=0.8cm}
\psframe[linewidth=2pt](1,2)(7,8)
\rput(1.3,8.2){$\pmb{0}$}
\rput(0.8,7.7){$\pmb{0}$}
\rput(1.8,8.2){$\pmb{1}$}
\rput(0.8,7.2){$\pmb{1}$}
\rput(2.3,8.2){$\pmb{2}$}
\rput(0.8,6.7){$\pmb{2}$}
\rput(6.8,8.2){$\pmb{2^k\!-\!1}$}
\rput(0.4,2.3){$\pmb{2^k\!-\!1}$}
\rput(4.2,8.2){$\scriptstyle\pmb{2^{k-1}}$}
\rput(3.8,8.9){$\scriptstyle\pmb{2^{k-1}-1}$}
\psline[linewidth=1pt,arrowsize=1pt 4]{->}(3.6,8.75)(3.6,8)
\rput(0.5,5){$\scriptstyle\pmb{2^{k-1}}$}
\rput(0,5.8){$\scriptstyle\pmb{2^{k-1}-1}$}
\psline[linewidth=1pt,arrowsize=1pt 4]{->}(0.4,5.6)(1,5.4)
\psframe[linewidth=2pt](3.6,4.9)(4.2,5.5)
\rput(4.8,4.3){\rotatebox{-30}{\Huge $\pmb{\ddots}$}}
\rput(4.7,6.1){\rotatebox{60}{\Huge $\pmb{\ddots}$}}
\rput(3,5.9){\rotatebox{150}{\Huge $\pmb{\ddots}$}}
\rput(3.2,4.3){\rotatebox{250}{\Huge $\pmb{\ddots}$}}
\rput(2.8,8.2){$\pmb{\dots}$}
\rput(5.3,8.2){$\pmb{\dots}$}
\rput(0.8,4){$\pmb{\vdots}$}
\rput(0.8,6.2){$\pmb{\vdots}$}
\rput(4,1.5){\bf (a)}
\rput(0,7.5){$\party_1$}
\rput(1.5,8.8){$\party_2$}
\psline[linewidth=2pt](1.5,8)(1.5,2.5)
\rput(1.25,5.2){\small \rotatebox{90}{\bf\small vertical level 1}}
\psline[linewidth=2pt](1,2.5)(6.5,2.5)
\rput(3.7,2.3){\small \rotatebox{0}{\bf\small horizontal level 1}}
\psline[linewidth=2pt](6.5,2)(6.5,7.5)
\rput(6.7,5.2){\small \rotatebox{90}{\bf\small vertical level 1}}
\psline[linewidth=2pt](7,7.5)(1.5,7.5)
\rput(4,7.75){\small \rotatebox{0}{\bf\small horizontal level 1}}
\psline[linewidth=1.5pt](2,7.5)(2,3)
\rput(1.75,5.2){\small \rotatebox{90}{\bf\small vertical level 2}}
\psline[linewidth=1.5pt](1.5,3)(6,3)
\rput(3.7,2.8){\small \rotatebox{0}{\bf\small horizontal level 2}}
\psline[linewidth=1.5pt](6,2.5)(6,7)
\rput(6.2,5.2){\small \rotatebox{90}{\bf\small vertical level 2}}
\psline[linewidth=1.5pt](6.5,7)(2,7)
\rput(4.1,7.25){\small \rotatebox{0}{\bf\small horizontal level 2}}
\psline[linewidth=1.5pt](2.5,7)(2.5,3.5)
\rput(2.25,5.2){\small \rotatebox{90}{\bf\small vertical level 3}}
\psline[linewidth=1.5pt](2,3.5)(5.5,3.5)
\rput(3.8,3.3){\small \rotatebox{0}{\footnotesize horizontal level 3}}
\psline[linewidth=1.5pt](5.5,3)(5.5,6.5)
\rput(5.7,4.8){\small \rotatebox{90}{\bf\small vertical level 3}}
\psline[linewidth=1.5pt](6,6.5)(2.5,6.5)
\rput(4.3,6.75){\small \rotatebox{0}{\footnotesize horizontal level 3}}
%
\psframe[linewidth=1pt](9,2)(15,8)
\rput(9.3,8.2){$\pmb{0}$}
\rput(8.8,7.7){$\pmb{0}$}
\rput(9.8,8.2){$\pmb{1}$}
\rput(8.8,7.2){$\pmb{1}$}
\rput(10.3,8.2){$\pmb{2}$}
\rput(8.8,6.7){$\pmb{2}$}
\rput(14.8,8.2){$\pmb{2^k\!-\!1}$}
\rput(8.4,2.3){$\pmb{2^k\!-\!1}$}
\rput(11.6,8.55){$\scriptstyle \pmb{2^{k-1}-1}$}
\psline[linewidth=1pt,arrowsize=1pt 4]{->}(11.6,8.4)(11.6,8)
\rput(8,5.8){$\scriptstyle \pmb{2^{k-1}-1}$}
\psline[linewidth=1pt,arrowsize=1pt 4]{->}(8.4,5.6)(9,5.4)
\rput(13.1,1.5){$\pmb{2^{k-1}\!\!+\!\!1}$}
\psline[linewidth=1pt,arrowsize=1pt 4]{->}(12.6,1.7)(12.4,2)
\rput(15.9,5.4){$\pmb{2^{k-1}\!\!+\!\!1}$}
\psline[linewidth=1pt,arrowsize=1pt 4]{->}(15.6,5.2)(15,4.5)
\rput(11,6.1){\sf A}
\psline[linewidth=1pt,linestyle=dashed,arrowsize=1pt 4]{->}(10.8,6.1)(9.2,6.1)
\rput(13,6.1){\sf C}
\psline[linewidth=1pt,linestyle=dashed,arrowsize=1pt 4]{->}(13.2,6.1)(14.8,6.1)
\rput(11,4.1){\sf B}
\psline[linewidth=1pt,linestyle=dashed,arrowsize=1pt 4]{->}(10.8,4.1)(9.2,4.1)
\rput(13,4.1){\sf D}
\psline[linewidth=1pt,linestyle=dashed,arrowsize=1pt 4]{->}(13.2,4.1)(14.8,4.1)
\rput(11,8.2){$\pmb{\dots}$}
\rput(13.3,8.2){$\pmb{\dots}$}
\rput(8.8,3.8){$\pmb{\vdots}$}
\rput(8.8,6.3){$\pmb{\vdots}$}
\rput(11,1.5){\bf (b)}
\psline[linewidth=1pt](9.5,8)(9.5,2.5)
\psline[linewidth=1pt](9,2.5)(14.5,2.5)
\psline[linewidth=1pt](14.5,2)(14.5,7.5)
\psline[linewidth=1pt](15,7.5)(9.5,7.5)
\psline[linewidth=1pt](10,7.5)(10,3)
\psline[linewidth=1pt](9.5,3)(14,3)
\psline[linewidth=1pt](14,2.5)(14,7)
\psline[linewidth=1pt](14.5,7)(10,7)
\psline[linewidth=1pt](10.5,7)(10.5,3.5)
\psline[linewidth=1pt](10,3.5)(13.5,3.5)
\psline[linewidth=1pt](13.5,3)(13.5,6.5)
\psline[linewidth=1pt](14,6.5)(10.5,6.5)
\psline[linewidth=3pt](11.8,7.5)(11.8,8)
\psline[linewidth=1pt,linestyle=dotted](9,5.6)(15,5.6)
\psline[linewidth=3pt](9,5.2)(9.5,5.2)
\psline[linewidth=1pt,linestyle=dotted](9.5,5.2)(15,5.2)
\psline[linewidth=3pt](12.2,2)(12.2,2.5)
\psline[linewidth=3pt](15,4.8)(14.5,4.8)
\psline[linewidth=1pt,linestyle=dotted](14.5,4.8)(9,4.8)
\psline[linewidth=1pt,linestyle=dotted](15,4.4)(9,4.4)
\end{pspicture}
\vspace*{-0.3in}
\caption{\label{fig:hless}\em Illustrations of the arguments in the proof of Theorem~\ref{thm:hless}. The
dotted lines in {\bf (b)} are shown for visual clarities only.}
\vspace*{-0.1in}
\end{figure}
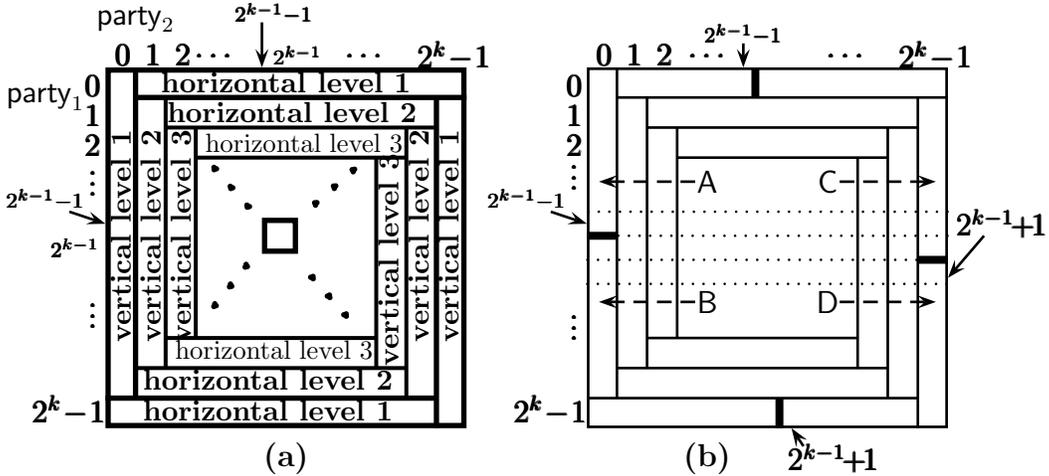

\noindent
{\bf (a)}
Consider the dissection protocol corresponding to the \BSP\ in Fact~\ref{bsp}.
Then, using $\max_{i}\{t_i\}\leq 4$ we get
$\alpha_{\scriptscriptstyle\UD^{\thicksim\,c}}\leq \sum_{i=1}^{\rr_f}\frac{4\,(1+c)\,y_i}{2^{2k}}=4\,(1+c)$.
Also, the number of communication steps in this protocol is the height of the \BSP-tree, which is at most $4\rr_f$.

\vspace*{0.1in}
\noindent
{\bf (b)}
Consider the function $f$
whose ideal monochromatic rectangles are shown
in Fig.~\ref{notile}.
Any correct protocol for computing $f$ {\em must} partition at least one rectangle of two elements,
giving
\[
\alpha_{\scriptscriptstyle\UD^{\thicksim\,c}}\geq 
4\times \left( \frac{1+c}{16}\right) + \frac{7-c}{8} 
=\frac{9+c}{8}
\]
\hfill{{\Pisymbol{pzd}{113}}}

\subsubsection{Large Worst-case \PAR\ for Non-Boolean Functions}

Can one extend the results of the last section to
show that every tiling function admits a dissection protocol that achieves a good \PAR\ {\em even in the worst case}?
We answer this question in the negative by
presenting a tiling function for which {\em every} dissection protocol has {\em large} worst-case \PAR.

\begin{Theorem}\label{thm:hless}
Let $k>0$ be an even integer. Then, there exists a tiling function $f:\{0,1\}^k\times\{0,1\}^k\mapsto \{0,1\}^3$ with respect to some two permutations $(\Pi_1,\Pi_2)$
such that, for any two permutations $\Pi_1'$ and $\Pi_2'$ of $\{0,1\}^k$,  every dissection protocol for $f$ with respect to
$(\Pi_1',\Pi_2')$ has $\aw>2^{k/2}-1$.
\end{Theorem}

\begin{proof}
Recall the example in Fig.~\ref{notile} that essentially showed that there exists functions that cannot be computed in a perfectly private manner.
Our construction of the function $f$ is based on the tiling shown in Fig.~\ref{notile}.
We consider the specific permutations $\Pi_1,\Pi_2$ over $\{0,1\}^k$ that order the elements in $\{0,1\}^k$ by binary value (from $0$ to $2^k-1$).
We now use the construction in Fig.~\ref{notile} ``recursively'' to create a tiling of the input space. We first embed $\frac{2^k-2}{2}=2^{k-1}-1$ instances of the
construction in Figure~\ref{notile} recursively within one another, as shown in Fig.~\ref{fig:hless}(a), leaving a $1\times 1$ square at the center.
The vertical level $i$ and the horizontal level $i$ rectangles have dimension $1\times \left(2^k-(2i-1)\right)$ and $\left(2^k-(2i-1)\right)\times 1$, respectively, for
$i=1,2,\dots,2^{k-1}-1$.
We then partition each of the level~1 rectangle in Fig.~\ref{fig:hless}(a) into two ``nearly'' equal-sized rectangles as shown in Fig.~\ref{fig:hless}(b).
Consider the function $f$ such that the monochromatic rectangles of $A_f(\Pi_1,\Pi_2)$
are the tilings in Fig.~\ref{fig:hless}(b) ($f$ outputs a different outcome for each (minimal) rectangle in the figure).
Clearly, $f$ is a tiling function with respect to $(\Pi_1,\Pi_2)$ and, moreover, since every rectangle shares a side with no more than
$8$ rectangles, at most $8$ output values of $f$ suffice.

Let $\Pi_1',\Pi_2'$ be any two arbitrary permutations of $\{0,1\}^k$ and consider any dissection protocol $P$
with respect to $(\Pi_1',\Pi_2')$. Consider the first meaningful step in the execution of $P$ and
suppose that this step was executed by $\party_1$ (the case that the step was executed by $\party_2$ is analogous).
This step partitions the total input space $\cS=\left\{0,1,2,\dots,2^k-1\right\}$ into two {\em nonempty} subsets,
say $I\subset\cS$ and $I'=\cS\setminus I$ such that $0\in I$. Let $0<i<2^k-1$ be the {\em least} integer such that
$i\in I$ but $i+1\not\in I$; such an $i$ must exist since both the sets are non-empty.
Consider the rectangles $A,B,C$ and $D$ in Fig.~\ref{fig:hless}(b). We have the following cases.

\vspace*{0.1in}
\noindent
{\bf Case 1:}
$i\leq 2^{k-1}-2^{k/2}$.
Observe that, for every such $i$, there exists a level $i+1$ vertical rectangle of size $2^k-2i-1$ that is
partitioned into two rectangles, one of which is of size exactly $1$.
Thus, $\aw\geq 2^k-2i+1>2^{(k/2)+1}-1>2^{k/2}-1$.

\vspace*{0.1in}
\noindent
{\bf Case 2:}
$2^{k-1}-2^{k/2}<i<2^{k-1}-1$.
Observe that, for every such value of $i$, rectangle $A$, which is of size $2^{k-1}$,
is partitioned into two rectangles, of which one is of size at most $2^{k/2}$. Thus, in this case
$\aw\geq\frac{2^k-1}{2^{k/2}}>2^{k/2}-1$.

\vspace*{0.1in}
\noindent
{\bf Case 3:}
$2^{k-1}-1\leq i\leq 2^{k-1}+1$.
In this case, at least one of the rectangles $B$, $C$ or $D$ is partitioned into two parts one of which
is of size at most $2$ and thus
$\aw\geq\frac{2^{k-1}-1}{2}>2^{k-2}-\frac{1}{2}>2^{k/2}-1$.

\vspace*{0.1in}
\noindent
{\bf Case 4:}
$2^{k-1}+1<i<2^k-2^{k/2}$.
Similar to Case~{\bf 2}.

\vspace*{0.1in}
\noindent
{\bf Case 5:}
$i\geq 2^k-2^{k/2}$.
Similar to Case~{\bf 1}.
\end{proof}

\section{Extensions of the Basic Two-party Setup}
\label{extensions}

\subsection{Non-tiling Functions}

A natural extension of the class of tiling functions involves relaxing the constraint
that each monochromatic region {\em must} be a rectangle.

\begin{Definition}[$\delta$-tiling function]
A function $f:\{0,1\}^k\times\{0,1\}^k\mapsto \{0,1\}^t$ is
a called a $\delta$-tiling function with respect to
permutations $(\Pi_1,\Pi_2)$ of $\{0,1\}^k$ if each maximal monochromatic region
of $A_{\Pi_1,\Pi_2}(f)$ is an union of at most $\delta$ disjoint rectangles.
\end{Definition}

For example, the function whose tiling is as shown in Fig.~\ref{some1} is a $2$-tiling Boolean function.

\begin{Proposition}\label{nontile}
For any $\delta$-tiling function $f$ with respect to $(\Pi_1,\Pi_2)$ with $r$ maximal monochromatic regions,
there is a dissection protocol $P$ with respect to $(\Pi_1,\Pi_2)$ using at most $4r\delta$ communication steps
such that
$\alpha_{\scriptscriptstyle\UD^{\thicksim\,c}}\leq (4+4c)\,\delta$.
\end{Proposition}

\begin{proof}
We use the algorithm of Theorem~\ref{x} on the set of at most $r\delta$ rectangles
obtained by partitioning each monochromatic region into rectangles.
Since each rectangle is partitioned at most $4$ times, each maximal monochromatic region of
$A_f(\Pi_1,\Pi_2)$ will be partitioned at most $4\delta$ times.
\end{proof}

\subsection{Multi-party Computation}

How good is the average \PAR\ for a dissection protocol on a $d$-dimensional tiling function? For a general $d$, it is
non-trivial to compute {\em precise} bounds because each $\party_i$ has her/his own permutation $\Pi_i$ of the
input, the tiles are boxes of {\em full} dimension and
hyperplanes corresponding to each step of the dissection protocol is of dimension {\em exactly} $d-1$.
Nonetheless, we show that the average \PAR\ is very high for dissection protocols
even for $3$ parties and uniform distribution, thereby suggesting that this quantification
of privacy may not provide good bounds for three or more parties.

\begin{figure}[htbp]
%
\begin{pspicture}(-0.3,-1.3)(7,5.5)
\psset{xunit=0.4cm,yunit=0.4cm}
\pspolygon[linewidth=0pt,fillstyle=solid,fillcolor=magenta](0,0)(0,8)(5.5,13.5)(13.5,13.5)(13.5,5.5)(8,0)
\pspolygon[linewidth=0pt,fillstyle=solid,fillcolor=darkgray,hatchsep=12pt,origin={-2,-2},swapaxes=true](3,3)(3.5,3.5)(11.5,3.5)(11,3)
\pspolygon[linewidth=0pt,fillstyle=solid,fillcolor=darkgray,hatchsep=12pt,origin={1,1},swapaxes=true](0,0.5)(0.5,1.1)(8.5,1.1)(8,0.5)
\pspolygon[linewidth=0pt,fillstyle=solid,fillcolor=darkgray,hatchsep=12pt,origin={1,1},swapaxes=true](0,0)(0,0.5)(8,0.5)(8,0)
\pspolygon[linewidth=0pt,fillstyle=solid,fillcolor=darkgray,hatchsep=12pt,origin={1,1},swapaxes=true](0.5,0.5)(0.5,1.1)(8.5,1.1)(8.5,0.5)
\pspolygon[linewidth=0pt,fillstyle=solid,fillcolor=darkgray,hatchsep=12pt,origin={2,-2},swapaxes=true](3,3)(3.5,3.5)(11.5,3.5)(11,3)
\pspolygon[linewidth=0pt,fillstyle=solid,fillcolor=darkgray,hatchsep=12pt,origin={5,1},swapaxes=true](0,0.5)(0.5,1.1)(8.5,1.1)(8,0.5)
\pspolygon[linewidth=0pt,fillstyle=solid,fillcolor=darkgray,hatchsep=12pt,origin={5,1},swapaxes=true](0,0)(0,0.5)(8,0.5)(8,0)
\pspolygon[linewidth=0pt,fillstyle=solid,fillcolor=darkgray,hatchsep=12pt,origin={5,1},swapaxes=true](0.5,0.5)(0.5,1.1)(8.5,1.1)(8.5,0.5)
\pspolygon[linewidth=0pt,fillstyle=solid,fillcolor=darkgray,hatchsep=12pt,origin={0,0},swapaxes=true](3,3)(3.5,3.5)(11.5,3.5)(11,3)
\pspolygon[linewidth=0pt,fillstyle=solid,fillcolor=darkgray,hatchsep=12pt,origin={3,3},swapaxes=true](0,0.5)(0.5,1.1)(8.5,1.1)(8,0.5)
\pspolygon[linewidth=0pt,fillstyle=solid,fillcolor=darkgray,hatchsep=12pt,origin={3,3},swapaxes=true](0,0)(0,0.5)(8,0.5)(8,0)
\pspolygon[linewidth=0pt,fillstyle=solid,fillcolor=darkgray,hatchsep=12pt,origin={3,3},swapaxes=true](0.5,0.5)(0.5,1.1)(8.5,1.1)(8.5,0.5)
\pspolygon[linewidth=0pt,fillstyle=solid,fillcolor=darkgray,hatchsep=12pt,origin={4,0},swapaxes=true](3,3)(3.5,3.5)(11.5,3.5)(11,3)
\pspolygon[linewidth=0pt,fillstyle=solid,fillcolor=darkgray,hatchsep=12pt,origin={7,3},swapaxes=true](0,0.5)(0.5,1.1)(8.5,1.1)(8,0.5)
\pspolygon[linewidth=0pt,fillstyle=solid,fillcolor=darkgray,hatchsep=12pt,origin={7,3},swapaxes=true](0,0)(0,0.5)(8,0.5)(8,0)
\pspolygon[linewidth=0pt,fillstyle=solid,fillcolor=darkgray,hatchsep=12pt,origin={7,3},swapaxes=true](0.5,0.5)(0.5,1.1)(8.5,1.1)(8.5,0.5)
\pspolygon[linewidth=0pt,fillstyle=solid,fillcolor=lightgray,hatchsep=6pt,origin={-0.5,-1}](0.5,3)(6,7.8)(6,8.8)(0.5,4)
\pspolygon[linewidth=0pt,fillstyle=solid,fillcolor=lightgray,hatchsep=6pt,origin={0.5,-1}](0.5,3)(6,7.8)(6,8.8)(0.5,4)
\pspolygon[linewidth=0pt,fillstyle=solid,fillcolor=lightgray,hatchsep=6pt,origin={-0.5,-1}](0.5,3)(1.5,3)(1.5,4)(0.5,4)
\pspolygon[linewidth=0pt,fillstyle=solid,fillcolor=lightgray,hatchsep=6pt,origin={5,3.8}](0.5,3)(1.5,3)(1.5,4)(0.5,4)
\pspolygon[linewidth=0pt,fillstyle=solid,fillcolor=lightgray,hatchsep=6pt,origin={3.5,-1}](0.5,3)(6,7.8)(6,8.8)(0.5,4)
\pspolygon[linewidth=0pt,fillstyle=solid,fillcolor=lightgray,hatchsep=6pt,origin={4.5,-1}](0.5,3)(6,7.8)(6,8.8)(0.5,4)
\pspolygon[linewidth=0pt,fillstyle=solid,fillcolor=lightgray,hatchsep=6pt,origin={3.5,-1}](0.5,3)(1.5,3)(1.5,4)(0.5,4)
\pspolygon[linewidth=0pt,fillstyle=solid,fillcolor=lightgray,hatchsep=6pt,origin={9,3.8}](0.5,3)(1.5,3)(1.5,4)(0.5,4)
\pspolygon[linewidth=0pt,fillstyle=solid,fillcolor=lightgray,hatchsep=6pt,origin={-0.5,3}](0.5,3)(6,7.8)(6,8.8)(0.5,4)
\pspolygon[linewidth=0pt,fillstyle=solid,fillcolor=lightgray,hatchsep=6pt,origin={0.5,3}](0.5,3)(6,7.8)(6,8.8)(0.5,4)
\pspolygon[linewidth=0pt,fillstyle=solid,fillcolor=lightgray,hatchsep=6pt,origin={-0.5,3}](0.5,3)(1.5,3)(1.5,4)(0.5,4)
\pspolygon[linewidth=0pt,fillstyle=solid,fillcolor=lightgray,hatchsep=6pt,origin={5,7.8}](0.5,3)(1.5,3)(1.5,4)(0.5,4)
\pspolygon[linewidth=0pt,fillstyle=solid,fillcolor=lightgray,hatchsep=6pt,origin={3.5,3}](0.5,3)(6,7.8)(6,8.8)(0.5,4)
\pspolygon[linewidth=0pt,fillstyle=solid,fillcolor=lightgray,hatchsep=6pt,origin={4.5,3}](0.5,3)(6,7.8)(6,8.8)(0.5,4)
\pspolygon[linewidth=0pt,fillstyle=solid,fillcolor=lightgray,hatchsep=6pt,origin={3.5,3}](0.5,3)(1.5,3)(1.5,4)(0.5,4)
\pspolygon[linewidth=0pt,fillstyle=solid,fillcolor=lightgray,hatchsep=6pt,origin={9,7.8}](0.5,3)(1.5,3)(1.5,4)(0.5,4)
\pspolygon[linewidth=1.5pt,fillstyle=solid,fillcolor=black,hatchsep=16pt](0,0)(0.5,0.5)(8.5,0.5)(8,0)
\pspolygon[linewidth=1.5pt,fillstyle=solid,fillcolor=black,hatchsep=16pt](0,0.5)(0.5,1.1)(8.5,1.1)(8,0.5)
\pspolygon[linewidth=1.5pt,fillstyle=solid,fillcolor=black,hatchsep=16pt](0,0)(0,0.5)(8,0.5)(8,0)
\pspolygon[linewidth=1.5pt,fillstyle=solid,fillcolor=black,hatchsep=16pt](0.5,0.5)(0.5,1.1)(8.5,1.1)(8.5,0.5)
\pspolygon[linewidth=1.5pt,fillstyle=solid,fillcolor=black,hatchsep=16pt,origin={0,4.3}](0,0)(0.5,0.5)(8.5,0.5)(8,0)
\pspolygon[linewidth=1.5pt,fillstyle=solid,fillcolor=black,hatchsep=16pt,origin={0,4.3}](0,0.5)(0.5,1.1)(8.5,1.1)(8,0.5)
\pspolygon[linewidth=1.5pt,fillstyle=solid,fillcolor=black,hatchsep=16pt,origin={0,4.3}](0,0)(0,0.5)(8,0.5)(8,0)
\pspolygon[linewidth=1.5pt,fillstyle=solid,fillcolor=black,hatchsep=16pt,origin={0,4.3}](0.5,0.5)(0.5,1.1)(8.5,1.1)(8.5,0.5)
\pspolygon[linewidth=1.5pt,fillstyle=solid,fillcolor=black,hatchsep=16pt](3,3)(3.5,3.5)(11.5,3.5)(11,3)
\pspolygon[linewidth=1.5pt,fillstyle=solid,fillcolor=black,hatchsep=16pt](3,3.5)(3.5,4.1)(11.5,4.1)(11,3.5)
\pspolygon[linewidth=1.5pt,fillstyle=solid,fillcolor=black,hatchsep=16pt](3,3)(3,3.5)(11,3.5)(11,3)
\pspolygon[linewidth=1.5pt,fillstyle=solid,fillcolor=black,hatchsep=16pt](3.5,3.5)(3.5,4.1)(11.5,4.1)(11.5,3.5)
\pspolygon[linewidth=1.5pt,fillstyle=solid,fillcolor=black,hatchsep=16pt,origin={0,4.3}](3,3)(3.5,3.5)(11.5,3.5)(11,3)
\pspolygon[linewidth=1.5pt,fillstyle=solid,fillcolor=black,hatchsep=16pt,origin={0,4.3}](3,3.5)(3.5,4.1)(11.5,4.1)(11,3.5)
\pspolygon[linewidth=1.5pt,fillstyle=solid,fillcolor=black,hatchsep=16pt,origin={0,4.3}](3,3)(3,3.5)(11,3.5)(11,3)
\pspolygon[linewidth=1.5pt,fillstyle=solid,fillcolor=black,hatchsep=16pt,origin={0,4.3}](3.5,3.5)(3.5,4.1)(11.5,4.1)(11.5,3.5)
%
%
\rput(3.5,-1.5){\bf (a)}
\hspace*{2.5in}
\pspolygon[linewidth=1.5pt,fillstyle=solid,fillcolor=black,hatchsep=16pt,xunit=0.9cm,origin={-2,0}](0,0)(0.5,0.5)(8.5,0.5)(8,0)
\pspolygon[linewidth=1.5pt,fillstyle=solid,fillcolor=black,hatchsep=16pt,xunit=0.9cm,origin={-2,0}](0,0.5)(0.5,1.1)(8.5,1.1)(8,0.5)
\pspolygon[linewidth=1.5pt,fillstyle=solid,fillcolor=black,hatchsep=16pt,xunit=0.9cm,origin={-2,0}](0,0)(0,0.5)(8,0.5)(8,0)
\pspolygon[linewidth=1.5pt,fillstyle=solid,fillcolor=black,hatchsep=16pt,xunit=0.9cm,origin={-2,0}](0.5,0.5)(0.5,1.1)(8.5,1.1)(8.5,0.5)
\pspolygon[linewidth=1.5pt,fillstyle=solid,fillcolor=black,hatchsep=16pt,origin={-2,4.3},xunit=0.9cm](0,0)(0.5,0.5)(8.5,0.5)(8,0)
\pspolygon[linewidth=1.5pt,fillstyle=solid,fillcolor=black,hatchsep=16pt,origin={-2,4.3},xunit=0.9cm](0,0.5)(0.5,1.1)(8.5,1.1)(8,0.5)
\pspolygon[linewidth=1.5pt,fillstyle=solid,fillcolor=black,hatchsep=16pt,origin={-2,4.3},xunit=0.9cm](0,0)(0,0.5)(8,0.5)(8,0)
\pspolygon[linewidth=1.5pt,fillstyle=solid,fillcolor=black,hatchsep=16pt,origin={-2,4.3},xunit=0.9cm](0.5,0.5)(0.5,1.1)(8.5,1.1)(8.5,0.5)
\pspolygon[linewidth=1.5pt,fillstyle=solid,fillcolor=black,hatchsep=16pt,xunit=0.9cm,origin={-2,0}](3,3)(3.5,3.5)(11.5,3.5)(11,3)
\pspolygon[linewidth=1.5pt,fillstyle=solid,fillcolor=black,hatchsep=16pt,xunit=0.9cm,origin={-2,0}](3,3.5)(3.5,4.1)(11.5,4.1)(11,3.5)
\pspolygon[linewidth=1.5pt,fillstyle=solid,fillcolor=black,hatchsep=16pt,xunit=0.9cm,origin={-2,0}](3,3)(3,3.5)(11,3.5)(11,3)
\pspolygon[linewidth=1.5pt,fillstyle=solid,fillcolor=black,hatchsep=16pt,xunit=0.9cm,origin={-2,0}](3.5,3.5)(3.5,4.1)(11.5,4.1)(11.5,3.5)
\pspolygon[linewidth=1.5pt,fillstyle=solid,fillcolor=black,hatchsep=16pt,origin={-2,4.3},xunit=0.9cm](3,3)(3.5,3.5)(11.5,3.5)(11,3)
\pspolygon[linewidth=1.5pt,fillstyle=solid,fillcolor=black,hatchsep=16pt,origin={-2,4.3},xunit=0.9cm](3,3.5)(3.5,4.1)(11.5,4.1)(11,3.5)
\pspolygon[linewidth=1.5pt,fillstyle=solid,fillcolor=black,hatchsep=16pt,origin={-2,4.3},xunit=0.9cm](3,3)(3,3.5)(11,3.5)(11,3)
\pspolygon[linewidth=1.5pt,fillstyle=solid,fillcolor=black,hatchsep=16pt,origin={-2,4.3},xunit=0.9cm](3.5,3.5)(3.5,4.1)(11.5,4.1)(11.5,3.5)
\psframe[linewidth=1.5pt,origin={5,11.5}](4,-2.5)(5,2.5)
\pspolygon[linewidth=0pt,fillstyle=gradient,gradbegin=lightgray,gradend=gray,origin={0,-2},swapaxes=true](-0.5,3)(11,9)(16,9)(4.5,3)
\pspolygon[linewidth=0pt,fillstyle=gradient,gradbegin=lightgray,gradend=gray,origin={1,-2},swapaxes=true](-0.5,3)(11,9)(16,9)(4.5,3)
\psframe[linewidth=1.5pt,origin={-1,0}](4,-2.5)(5,2.5)
\psframe[linewidth=1.5pt,origin={9,11.5}](4,-2.5)(5,2.5)
\pspolygon[linewidth=0pt,fillstyle=gradient,gradbegin=lightgray,gradend=gray,origin={4,-2},swapaxes=true](-0.5,3)(11,9)(16,9)(4.5,3)
\pspolygon[linewidth=0pt,fillstyle=gradient,gradbegin=lightgray,gradend=gray,origin={5,-2},swapaxes=true](-0.5,3)(11,9)(16,9)(4.5,3)
\psframe[linewidth=1.5pt,origin={3,0}](4,-2.5)(5,2.5)
\rput(2,-1.5){\bf (b)}
\psline[linewidth=2pt,arrowsize=1.5pt 4]{->}(-1.8,10)(-1.8,12)
\rput(0.8,12.5){$\scriptstyle\pmb{y}\!\!$ {\small (dimension $\scriptstyle 2$)}}
\psline[linewidth=2pt,arrowsize=1.5pt 4]{->}(-1.8,10)(0.2,10)
\rput(3.3,10){$\scriptstyle\pmb{x}\!\!$ {\small (dimension $\scriptstyle 1$)}}
\psline[linewidth=2pt,arrowsize=1.5pt 4]{->}(-1.8,10)(0.2,12)
\rput(3.2,11.7){$\scriptstyle\pmb{z}\!\!$ {\small (dimension $\scriptstyle 3$)}}
\end{pspicture}
\vspace*{-0.2in}
\caption{\label{3d}\em {\bf (not drawn to scale)} {\bf (a)} The tiling function in the proof of Lemma~\ref{y}. The non-trivial rectangles for dimensions $1$, $2$ and $3$
are colored by black, dark gray and light gray, respectively; the trivial rectangles, each having a distinct value, cover the region colored magenta.
{\bf (b)} Rectangles (in light gray) corresponding to a hypothetically first meaningful step of the protocol.}
\vspace*{-0.2in}
\end{figure}
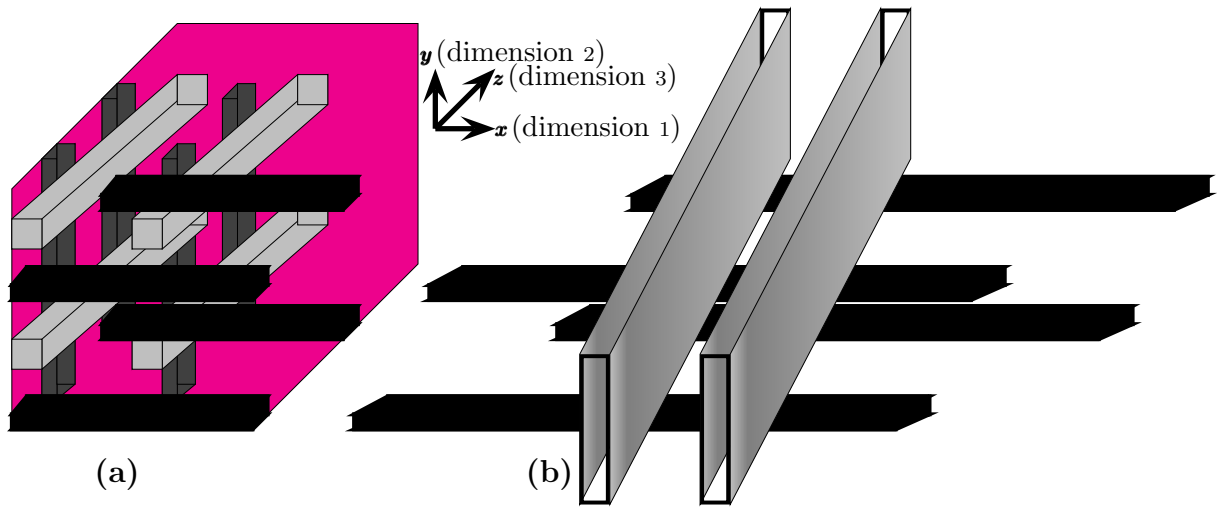

\begin{Theorem}[large average \PAR\ for dissection protocols with $3$ parties]\label{y}
There exists a tiling function
$f\colon \!{\{0,1\}^k\times\{0,1\}^k\times\{0,1\}^k}\!\mapsto\! \{0,1\}^{3k}$
such that, for any three permutations $\Pi_1,\Pi_2,\Pi_3$ of $\{0,1\}^k$, every dissection protocol with respect to $(\Pi_1,\Pi_2,\Pi_3)$ must have
$\alpha_{\scriptscriptstyle\UD}=\Omega\left(2^k\right)$.
\end{Theorem}

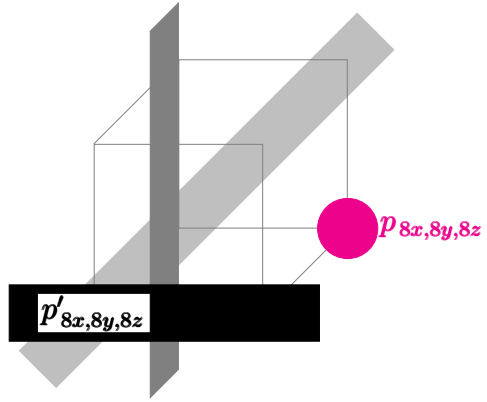
\begin{wrapfigure}[13]{R}{5.5cm}
\begin{pspicture}(-1,-2.5)(4,3.7)
\psset{xunit=0.75cm,yunit=0.75cm}
\psline[linewidth=20pt,linecolor=lightgray,fillstyle=solid,fillcolor=lightgray,origin={3,3}](-4,-4)(2,2)
\psframe[linewidth=0.3pt,linecolor=gray](0,0)(3,3)
\psframe[linewidth=0.3pt,linecolor=gray,origin={1.5,1.5}](0,0)(3,3)
\psline[linewidth=0.3pt,linecolor=gray,origin={3,0}](0,0)(1.5,1.5)
\psline[linewidth=0.3pt,linecolor=gray,origin={0,3}](0,0)(1.5,1.5)
\pscircle[linewidth=0pt,linecolor=magenta,fillstyle=solid,fillcolor=magenta,origin={4.5,1.5}](0,0){0.4}
\pspolygon[linewidth=0pt,linecolor=gray,fillstyle=solid,fillcolor=gray,origin={1.5,1}](-0.5,-2.5)(0,-2)(0,4.5)(-0.5,4)
\rput(6,1.5){\color{magenta}$\pmb{p_{\,8x,8y,8z}}$}
\psframe[linewidth=0pt,linecolor=black,fillstyle=solid,fillcolor=black](-1.5,-0.5)(4,0.5)
\psframe[linewidth=0pt,linecolor=black,fillstyle=solid,fillcolor=white,origin={0,0}](-1,-0.35)(1,0.35)
\rput(0,0){\color{black}$\pmb{p_{\,8x,8y,8z}'}$}
\end{pspicture}
\vspace*{-0.7in}
\caption{\label{sepr}\em Separating $p_{8x,8y,8z}$ from $p_{8x,8y,8z}'$.}
\end{wrapfigure}

\noindent
{\em Proof.}
In the sequel, for convenience {\em we refer to $3$-dimensional hyper-rectangles simply by rectangles}
and refer to the arguments of function $f$ via {\em decimal equivalent of the corresponding binary numbers}.
The tiling function for this theorem is adopted from an example of the paper by Paterson and Yao~\cite{paterson90,paterson92}
with appropriate modifications. The three arguments of $f$ are referred to as dimensions $1$, $2$ and $3$, respectively.
Define the {\em volume} of a rectangle $R=[x_1,x_1']\times [x_2,x_2']\times [x_3,x_3']\subseteq \{0,1,\dots,2^k-1\}^3$ is
$\vol(R)=\max\{0,\Pi_{i=1}^3 (x_i'-x_i+1)\}$.
For convenience, let $[\ast]$ denote the interval $\left[0,2^k-1\right]$.
We provide the tiling for the function $f$; see Fig.~\ref{3d} for a graphical illustration (note that Fig.~\ref{3d} is {\em not} drawn to scale):
\begin{itemize}
\item
For each dimension, we have a set of $\Theta\left(2^{2k}\right)$ rectangles; we refer to these rectangles as {\em non-trivial} rectangles for this dimension.
\begin{itemize}
\item
For dimension $1$, these rectangles are of the form $[\ast]\times[2y,2y]\times[2z,2z]$ for every integral value of $0\leq 2y,2z<2^k$.

\item
For dimension $2$, these rectangles are of the form $[2x,2x]\times[\ast]\times[2z+1,2z+1]$ for every integral value of $0\leq 2x,2z+1<2^k$.

\item
For dimension $3$, these rectangles are of the form $[2x+1,2x+1]\times[2y+1,2y+1]\times[\ast]$ for every integral value of $0\leq 2x+1,2y+1<2^k$.
\end{itemize}

\item
The remaining ``trivial'' rectangles are each of unit volume such that they together cover the remaining input space.
\end{itemize}
Let $\cSn$ be the set of all non-trivial rectangles. Observe that:
\begin{itemize}
\item
Rectangles in $\cSn$ are mutually disjoint since any two of them do not intersect in at least one dimension.

\item
{\em Each} rectangle in $\cSn$ has a volume of $2^k$ and thus the sum of their volumes is $\Theta\left(2^{3k}\right)$.
\end{itemize}
It now also follows that the number of monochromatic regions is $O\left(2^{3k}\right)$.
Suppose that a dissection protocol partitions, for $i=1,2,\dots,|\cSn|$,
the $i\tx$ non-trivial rectangle $R_i\in\cSn$ into $t_i$ rectangles, say $R_{i,1},R_{i,2},\dots,R_{i,t_i}$. Then,
\begin{gather*}
\alpha_{\scriptscriptstyle\UD}
\!\!
\stackrel{\mathrm{\scriptstyle def}}{=}
\!\!\!\!\!\!\!\!\!\!\!\!\!\!\!\!\!\!\!\!
\sum_{\substack{(x,y,z)\,\in \\ \{0,1\}^k\times\{0,1\}^k\times\{0,1\}^k}}
\!\!\!\!\!\!\!\!\!\!\!\!\!\!\!\!\!\!\!\Pr_{\UD}\left[x\,\&\,y\,\&\,z\right]\frac{\left|R^I(x,y,z)\right|}{\left|R^P(x,y,z)\right|}
\geq
\sum_{i=1}^{|\cSn|} \sum_{j=1}^{t_i} \!\!\!\!\!\!\!\!\!\!\!\!\!\!
\sum_{\,\,\,\,\,\,\,\,\,\,\,\,\,\,\,\,\,\,(x,y,z)\in R_{i,j}}\!\!\!\!\!\!\!\!\!\!\!\!\!\!\!
\Pr_{\UD}[x\,\&\,y\,\&\,z]\,\frac{\vol\left(R_i\right)}{\vol\left(R_{i,j}\right)}
\\
=
\sum_{i=1}^{|\cSn|} \sum_{j=1}^{t_i} \frac{2^k}{2^{3k}}
=
\sum_{i=1}^{|\cSn|}\!\!\!\!\left(t_i/2^{2k}\right)
\end{gather*}
Thus, it suffices to show that $\displaystyle\sum_{i=1}^{|\cSn|}\!\!\!\!\!\!\!\!t_i=\Omega\left(2^{3k}\right)$.
Let $\Q$ be the set of maximal {\em monochromatic} rectangles produced the partitioning of the entire protocol.
Consider the two entries $p_{x,y,z}=(2x+1,2y,2z+1)$ and $p_{x,y,z}'=(2x,2y,2z)$ (see Fig.~\ref{sepr}).
Note that $p_{x,y,z}$ belongs to a trivial rectangle
since their third, first and second coordinate does not lie within {\em any} non-trivial rectangle of dimension $1$, $2$ and
$3$, respectively, whereas $p_{x,y,z}'$ belongs to the non-trivial rectangle $[\ast]\times[2\times (8y),2\times (8y)]\times[2\times (8z),2\times (8z)]$ of
dimension~$1$. Thus, $p_{x,y,z}$ and $p_{x,y,z}'$ cannot belong to the same rectangle in $\Q$.
Let $T=\bigcup \big\{\,\{p_{\,8x,8y,8z},\,p_{\,8x,8y,8z}'\}\,|\,64<16x,16y,16z<2^k-64\,\big\}$.
Clearly, $|T|=\Theta\left(2^{3k}\right)$.
For an entry $(x_1,x_2,x_3)$, let its neighborhood be defined by the ball
$\nbr(x_1,x_2,x_3)=\big\{\,(x_1',x_2',x_3')\,|\,\forall i\,\colon |x_i-x_i'|\leq 4\,\big\}$, \IE, 
the neighborhood of an entry is the set of all entries $(x_1',x_2',x_3')$ such that 
each $x_i'$ lies in the range $\big[x_i-4,x_i+4\big]$ for $i=1,2,3$.
Note that $\nbr(p_{\,8x,8y,8z})\cap\nbr(p_{\,8x',8y',8z'})=\emptyset$ provided $(x,y,z)\neq (x',y',z')$.
Next, we show that, to ensure that the two entries $p_{\,8x,8y,8z}$ and $p_{\,8x,8y,8z}'$
are in two different rectangles in $\Q$,
the protocol must produce an {\em additional} fragment of one of the non-trivial rectangles in the neighborhood
$\nbr(p_{\,8x,8y,8z})$; this would directly imply $\sum_i t_i=\Omega\left(2^{3k}\right)$.

Consider the step of the protocol {\em before} which
$p_{\,8x,8y,8z}$ and $p_{\,8x,8y,8z}'$ were contained inside the same rectangle, namely a
rectangle $Q$ that includes the rectangle $[16x,16x+1]\times [16y,16y]\times [16z,16z+1]$, but after which they are in two different
rectangles $Q_1=[a_1',b_1']\times[a_2',b_2']\times[a_3',b_3']$ and $Q_2=[a_1'',b_1'']\times[a_2'',b_2'']\times[a_3'',b_3'']$.
Remember that both $Q_1$ and $Q_2$ must have the same two dimensions and these two dimensions must be the same as the corresponding
dimensions of $Q$. The following cases arise.

\vspace*{0.1in}
\noindent
{\bf Case~1 (split via the first coordinate):}
$[a_2',b_2']=[a_2'',b_2'']\supseteq [16y,16y]$, $[a_3',b_3']=[a_3'',b_3'']\supseteq [16z,16z+1]$,
$b_1'=16x$ and $a_1''=16x+1$.
Then, a new fragment of a non-trivial rectangle of dimension~$2$ is produced at
$[16x,16y,16z]\in\nbr(p_{\,8x,8y,8z})$.

\vspace*{0.2in}
\noindent
{\bf Case~2 (split via the second coordinate):}
$[a_1',b_1']=[a_1'',b_1'']\supseteq [16x,16x+1]$ and $[a_3',b_3']=[a_3'',b_3'']\supseteq [16z,16z+1]$.
This case is not possible.

\vspace*{0.2in}
\noindent
{\bf Case~3 (split via the third coordinate):}
$[a_1',b_1']=[a_1'',b_1'']\supseteq [16x,16x+1]$, $[a_2',b_2']=[a_2'',b_2'']\supseteq [16y,16y]$,
$b_3'=16z$ and $a_3''=16z+1$.
Then, a new fragment of a non-trivial rectangle of dimension~$1$ is produced at
$[16x,16y,16z]\in\nbr(p_{\,8x,8y,8z})$.
{\hfill{\Pisymbol{pzd}{113}}\vspace{0.1in}}

\begin{remark}
A generalized version of the example in $d$ dimension can be used to
provide a slightly improved lower bound on $\alpha_{\scriptscriptstyle\UD}$ for dissection protocols with more than three parties;
the bound asymptotically approaches $\Omega\left(2^{2k}\right)$ for large $d$.
\end{remark}

\section{Analysis of the Bisection Protocol for Two Functions}
\label{two-f}

In Section~\ref{bool} we showed that any Boolean tiling function can be computed with perfect privacy by a dissection protocol.
In~\cite{FJS10} the authors provided calculated bounds on $\aw$ and
$\alpha_{\scriptscriptstyle\UD}$ for the {\em bisection} protocol, a special case of the general dissection protocol (see Definition~\ref{bidef}),
on a few functions. In this section, we analyze the bisection protocol~\cite{muller,muller2},
for two Boolean functions that appear in the literature.
As before, $\UD$ denotes the uniform distribution.
Letting $\x=(x_1,x_2,\ldots,x_n)\in\{0,1\}^k$ and $\y=(y_1,y_2,\ldots,y_n)\in \{0,1\}^k$,
the functions that we consider are the following:
\begin{description}
\item[{\sf set-covering}:]
$f_{\wedge,\vee}(\x,\y)=\bigwedge_{i=1}^n\left(x_i\vee y_i\right)$.
To interpret this as a set-covering function, suppose that the universe $\cU$ consists of $n$ elements $e_1,e_2,\dots,e_n$
and the vectors $\x$ and $\y$ encode membership of the elements in two sets $S_{\x}$ and $S_{\y}$, \IE,
$x_i$ (respectively, $y_i$) is $1$ if and only if $e_i\in S_{\x}$ (respectively, $e_i\in S_{\y}$).
Then, $f_{\wedge,\vee}(\x,\y)=1$ if and only if $S_{\x}\cup S_{\y}=\cU$.

\item[{\sf equality}:]
$f_{=}(\x,\y)=\left\{\begin{array}{ll}1 & \mbox{if $\forall\,i:\,x_i=y_i$} \\ 0 & \mbox{otherwise} \\\end{array}\right.$.
The equality function provides a useful testbed for evaluating privacy preserving protocols, \EG, see~\cite{BCKO93}.
\end{description}
As we already noted in Section~\ref{contribution}, both of these functions are studied in the context of
evaluating privacy preserving protocols and communication complexity settings~\cite{BCKO93,HN97}.
A summary of our bounds are as follows.

\begin{center}
\begin{tabular}{c|cc} \toprule
$f_{\wedge,\vee}\,\,\,\,$ & \multicolumn{2}{c}{$\aw\geq \alpha_{\scriptscriptstyle\UD}\geq\left(\frac{3}{2}\right)^{2k}$} \\ \midrule
$f_{=}\,\,\,\,$ & $\,\,\alpha_{\scriptscriptstyle\UD}\!\!=2^k-2+2^{1-k}\,\,\,\,\,\,$ & $\aw=2^{2k-1}-2^{k-1}$ \\
\bottomrule
\end{tabular}
\end{center}

We will use the formula for $\alpha_{\scriptscriptstyle\UD}$ that we derived in the
proof of Theorem~\ref{x}: {\em letting $r$ denote the number of monochromatic regions in an ideal partition of
the function if, for $i=1,2,\dots,r$,
the $i\tx$ monochromatic region contain $y_i\times 2^{2k}$ elements
and the bisection protocol partitions this region
into $t_i\geq 1$ rectangles containing $z_1,\dots,z_{t_i}$ elements,
respectively, then $\alpha_{\scriptscriptstyle\UD}=\sum_{i=1}^rt_iy_i$}.
In the sequel, by ``contribution of a rectangle (of the bisection protocol) to the
(average \PAR)'' we mean the size of the ideal monochromatic region that the rectangle is a part.

\subsection{Set Covering Function}

\begin{Theorem}
$\alpha_{\scriptscriptstyle\UD}\geq\left(3/2\right)^{2k}$.
\end{Theorem}

\begin{proof}
We begin by showing the geometry of the tilings for small values of $k$
which easily generalizes to larger $k$.
The ideal tiling for $f_{\wedge,\vee}$ is shown in Fig.~\ref{f12}(a) for $k=3$
with the value of the function for each input pair.
The sizes of the ideal monochromatic partition are shown in Fig.~\ref{f12}(b) for $k=1,2,3,4$.
The contributions to the average \PAR\ of various inputs
after applying the bisection protocol are illustrated in Fig.~\ref{f3} for $k=1,2,3,4$.
We observe the following:
\begin{itemize}
\item
The tiles colored {\em light gray} for the case when $k=4$ are referred to as the ``background tiles''.
For $k=1,2,3,4$ each such tile contributes $3,9,27$ and $81$, respectively, to the average \PAR.
In general, this contribution is given by $3^k$ and all these tiles have size $1$.

\item
The contributions of the tiles in the upper-left region of the matrix are given by the sum of the first
$2^k-1$ natural numbers; thus each of these tiles contribute $2^{2k-1}-2^{k-1}$.

\item
For any $k$, observe that the matrix can be decomposed into $4$ quadrants;
the following observations can be repeated recursively on each resulting quadrant, except for the first quadrant:
\begin{itemize}
\item
The first quadrant is a monochromatic region that contributes $2^{2k-1}-2^{k-1}$ to the average \PAR.

\item
The fourth quadrant has the same structure as the original matrix, but the contributions for the
{\em non-background} tiles will be related to the case of a matrix with $j$ bits instead of $k$, where the size of the quadrant is
$2^j$. For example, notice that the fourth quadrant of a matrix with $k=4$ is the same as a whole matrix with $k=3$,
except for the ``background tiles'', that always contribute for $3^k$, with the original value of $k$.

\item
The second and third quadrants are similar to the fourth quadrant case, but in this case the values in the
upper-left portion of the quadrants will remain the same as the original matrix, instead of going down as with the
fourth quadrant case.
\end{itemize}
\end{itemize}

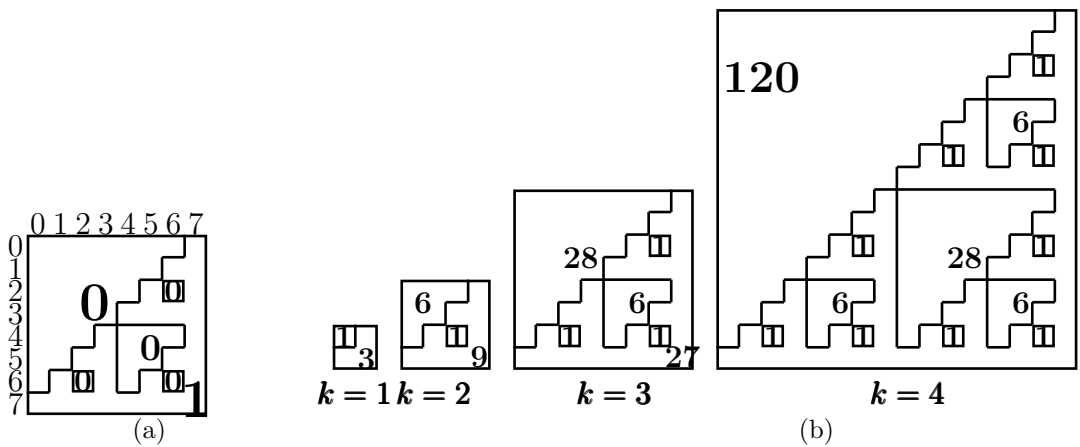
\begin{figure}[h]
\hspace*{-0.3in}
\subfigure[]
{
\begin{pspicture}(0,0.5)(4,2)
\psset{xunit=0.3cm,yunit=0.3cm}
\psframe[linewidth=1pt](1,1)(9,9)
\psline[linewidth=1pt](1,2)(2,2)
\psline[linewidth=1pt](2,2)(2,3)
\psline[linewidth=1pt](2,3)(3,3)
\psline[linewidth=1pt](3,3)(3,4)
\psline[linewidth=1pt](3,4)(4,4)
\psline[linewidth=1pt](4,4)(4,5)
\psline[linewidth=1pt](4,5)(5,5)
\psline[linewidth=1pt](5,5)(5,6)
\psline[linewidth=1pt](5,6)(6,6)
\psline[linewidth=1pt](6,6)(6,7)
\psline[linewidth=1pt](6,7)(7,7)
\psline[linewidth=1pt](7,7)(7,8)
\psline[linewidth=1pt](7,8)(8,8)
\psline[linewidth=1pt](8,8)(8,9)
\rput(4,6){{\LARGE\bf 0}}
\rput(8.5,1.7){{\LARGE\bf 1}}
\psframe[linewidth=1pt](4,2)(3,3)
\rput(3.5,2.5){{\bf 0}}
\psline[linewidth=1pt](5,5)(5,2)
\psline[linewidth=1pt](5,2)(6,2)
\psline[linewidth=1pt](6,2)(6,3)
\psline[linewidth=1pt](6,3)(7,3)
\psframe[linewidth=1pt](7,3)(8,2)
\rput(7.5,2.5){{\bf 0}}
\psline[linewidth=1pt](7,3)(7,4)
\psline[linewidth=1pt](7,4)(8,4)
\psline[linewidth=1pt](8,4)(8,5)
\psline[linewidth=1pt](8,5)(5,5)
\psframe[linewidth=1pt](7,7)(8,6)
\rput(7.5,6.5){{\bf 0}}
\rput(6.5,4){{\bf\large 0}}
\rput(0.5,8.5){0}
\rput(0.5,7.5){1}
\rput(0.5,6.5){2}
\rput(0.5,5.5){3}
\rput(0.5,4.5){4}
\rput(0.5,3.5){5}
\rput(0.5,2.5){6}
\rput(0.5,1.5){7}
\rput(1.5,9.5){0}
\rput(2.5,9.5){1}
\rput(3.5,9.5){2}
\rput(4.5,9.5){3}
\rput(5.5,9.5){4}
\rput(6.5,9.5){5}
\rput(7.5,9.5){6}
\rput(8.5,9.5){7}
\end{pspicture}
}
\subfigure[]
{
\begin{pspicture}(0,-0.4)(13,5)
\psset{xunit=0.3cm,yunit=0.3cm}
%
\psframe[linewidth=1pt](0,0)(2,2)
\psline[linewidth=1pt](0,1)(1,1)
\psline[linewidth=1pt](1,1)(1,2)
\rput(0.5,1.5){{\bf 1}}
\rput(1.5,0.5){{\bf 3}}
\rput(1,-1){{$\pmb{k=1}$}}
%
\psframe[linewidth=1pt](3,0)(7,4)
\psline[linewidth=1pt](3,1)(4,1)
\psline[linewidth=1pt](4,1)(4,2)
\psline[linewidth=1pt](4,2)(5,2)
\psline[linewidth=1pt](5,2)(5,3)
\psline[linewidth=1pt](5,3)(6,3)
\psline[linewidth=1pt](6,3)(6,4)
\psframe[linewidth=1pt](5,2)(6,1)
\rput(5.5,1.5){{\bf 1}}
\rput(4,3){{\bf 6}}
\rput(6.5,0.6){{\bf 9}}
\rput(4.5,-1){{$\pmb{k=2}$}}
\rput(12.5,-1){{$\pmb{k=3}$}}
\psframe[linewidth=1pt](8,0)(16,8)
\psline[linewidth=1pt](8,1)(9,1)
\psline[linewidth=1pt](9,1)(9,2)
\psline[linewidth=1pt](9,2)(10,2)
\psline[linewidth=1pt](10,2)(10,3)
\psline[linewidth=1pt](10,3)(11,3)
\psline[linewidth=1pt](11,3)(11,4)
\psline[linewidth=1pt](11,4)(12,4)
\psline[linewidth=1pt](12,4)(12,5)
\psline[linewidth=1pt](12,5)(13,5)
\psline[linewidth=1pt](13,5)(13,6)
\psline[linewidth=1pt](13,6)(14,6)
\psline[linewidth=1pt](14,6)(14,7)
\psline[linewidth=1pt](14,7)(15,7)
\psline[linewidth=1pt](15,7)(15,8)
\rput(11,5){{\bf 28}}
\rput(15.5,0.6){{\bf 27}}
\psframe[linewidth=1pt](11,1)(10,2)
\rput(10.5,1.5){{\bf 1}}
\psline[linewidth=1pt](12,4)(12,1)
\psline[linewidth=1pt](12,1)(13,1)
\psline[linewidth=1pt](13,1)(13,2)
\psline[linewidth=1pt](13,2)(14,2)
\psframe[linewidth=1pt](14,2)(15,1)
\rput(14.5,1.5){{\bf 1}}
\psline[linewidth=1pt](14,2)(14,3)
\psline[linewidth=1pt](14,3)(15,3)
\psline[linewidth=1pt](15,3)(15,4)
\psline[linewidth=1pt](15,4)(12,4)
\psframe[linewidth=1pt](14,6)(15,5)
\rput(14.5,5.5){{\bf 1}}
\rput(13.5,3){{\bf 6}}
\psset{origin={9,0}}
\psframe[linewidth=1pt](8,0)(24,16)
\rput(19,13){{\Large\bf 120}}
\rput(25.5,-1){{$\pmb{k=4}$}}
\psline[linewidth=1pt](8,1)(9,1)
\psline[linewidth=1pt](9,1)(9,2)
\psline[linewidth=1pt](9,2)(10,2)
\psline[linewidth=1pt](10,2)(10,3)
\psline[linewidth=1pt](10,3)(11,3)
\psline[linewidth=1pt](11,3)(11,4)
\psline[linewidth=1pt](11,4)(12,4)
\psline[linewidth=1pt](12,4)(12,5)
\psline[linewidth=1pt](12,5)(13,5)
\psline[linewidth=1pt](13,5)(13,6)
\psline[linewidth=1pt](13,6)(14,6)
\psline[linewidth=1pt](14,6)(14,7)
\psline[linewidth=1pt](14,7)(15,7)
\psline[linewidth=1pt](15,7)(15,8)
\psline[linewidth=1pt](15,8)(16,8)
\psframe[linewidth=1pt](11,1)(10,2)
\rput(19.5,1.5){{\bf 1}}
\psline[linewidth=1pt](12,4)(12,1)
\psline[linewidth=1pt](12,1)(13,1)
\psline[linewidth=1pt](13,1)(13,2)
\psline[linewidth=1pt](13,2)(14,2)
\psframe[linewidth=1pt](14,2)(15,1)
\rput(23.5,1.5){{\bf 1}}
\psline[linewidth=1pt](14,2)(14,3)
\psline[linewidth=1pt](14,3)(15,3)
\psline[linewidth=1pt](15,3)(15,4)
\psline[linewidth=1pt](15,4)(12,4)
\psframe[linewidth=1pt](14,6)(15,5)
\rput(23.5,5.5){{\bf 1}}
\rput(22.5,3){{\bf 6}}
\psset{origin={17,0}}
\psline[linewidth=1pt](8,1)(9,1)
\psline[linewidth=1pt](9,1)(9,2)
\psline[linewidth=1pt](9,2)(10,2)
\psline[linewidth=1pt](10,2)(10,3)
\psline[linewidth=1pt](10,3)(11,3)
\psline[linewidth=1pt](11,3)(11,4)
\psline[linewidth=1pt](11,4)(12,4)
\psline[linewidth=1pt](12,4)(12,5)
\psline[linewidth=1pt](12,5)(13,5)
\psline[linewidth=1pt](13,5)(13,6)
\psline[linewidth=1pt](13,6)(14,6)
\psline[linewidth=1pt](14,6)(14,7)
\psline[linewidth=1pt](14,7)(15,7)
\psline[linewidth=1pt](15,7)(15,8)
\rput(28,5){{\bf 28}}
\psframe[linewidth=1pt](11,1)(10,2)
\rput(27.5,1.5){{\bf 1}}
\psline[linewidth=1pt](12,4)(12,1)
\psline[linewidth=1pt](12,1)(13,1)
\psline[linewidth=1pt](13,1)(13,2)
\psline[linewidth=1pt](13,2)(14,2)
\psframe[linewidth=1pt](14,2)(15,1)
\rput(31.5,1.5){{\bf 1}}
\psline[linewidth=1pt](14,2)(14,3)
\psline[linewidth=1pt](14,3)(15,3)
\psline[linewidth=1pt](15,3)(15,4)
\psline[linewidth=1pt](15,4)(12,4)
\psframe[linewidth=1pt](14,6)(15,5)
\rput(31.5,5.5){{\bf 1}}
\rput(30.5,3){{\bf 6}}
\psset{origin={17,8}}
\psline[linewidth=1pt](8,1)(9,1)
\psline[linewidth=1pt](9,1)(9,2)
\psline[linewidth=1pt](9,2)(10,2)
\psline[linewidth=1pt](10,2)(10,3)
\psline[linewidth=1pt](10,3)(11,3)
\psline[linewidth=1pt](11,3)(11,4)
\psline[linewidth=1pt](11,4)(12,4)
\psline[linewidth=1pt](12,4)(12,5)
\psline[linewidth=1pt](12,5)(13,5)
\psline[linewidth=1pt](13,5)(13,6)
\psline[linewidth=1pt](13,6)(14,6)
\psline[linewidth=1pt](14,6)(14,7)
\psline[linewidth=1pt](14,7)(15,7)
\psline[linewidth=1pt](15,7)(15,8)
\psframe[linewidth=1pt](11,1)(10,2)
\rput(27.5,9.5){{\bf 1}}
\psline[linewidth=1pt](12,4)(12,1)
\psline[linewidth=1pt](12,1)(13,1)
\psline[linewidth=1pt](13,1)(13,2)
\psline[linewidth=1pt](13,2)(14,2)
\psframe[linewidth=1pt](14,2)(15,1)
\rput(31.5,9.5){{\bf 1}}
\psline[linewidth=1pt](14,2)(14,3)
\psline[linewidth=1pt](14,3)(15,3)
\psline[linewidth=1pt](15,3)(15,4)
\psline[linewidth=1pt](15,4)(12,4)
\psframe[linewidth=1pt](14,6)(15,5)
\rput(31.5,13.5){{\bf 1}}
\rput(30.5,11){{\bf 6}}
\psline[linewidth=1pt](8,1)(8,-7)
\psline[linewidth=1pt](8,0)(15,0)
\end{pspicture}
}
\caption{\label{f12}{\bf (a)} Ideal monochromatic partition for $f_{\wedge,\vee}$ when $k=3$.
{\bf (b)} Sizes of ideal monochromatic partition for $f_{\wedge,\vee}$.}
\end{figure}

\begin{figure}[htbp]
\hspace*{0.25in}
\subtable[$\pmb{k=1}$]{
\begin{tabular}{|c|c|} \hline
{\bf 1} & {\bf 3} \\ \hline
{\bf 3} & {\bf 3} \\ \hline
\end{tabular}
}
\subtable[$\pmb{k=2}$]{
\begin{tabular}{|c|c|c|c|} \hline
\multicolumn{2}{|c|}{} & {\bf 6} & {\bf 9} \\ \cline{3-4}
\multicolumn{2}{|c|}{{{\bf 6}}} & {\bf 9} & {\bf 9} \\ \hline
{\bf 6} & {\bf 9} & {\bf 1} & {\bf 9} \\ \hline
{\bf 9} & {\bf 9} & {\bf 9} & {\bf 9} \\ \hline
\end{tabular}
}
\subtable[$\pmb{k=3}$]{
\begin{tabular}{|c|c|c|c|c|c|c|c|} \hline
\multicolumn{4}{|c|}{} & \multicolumn{2}{|c|}{} & {\bf 28} & {\bf 27} \\ \cline{7-8}
\multicolumn{4}{|c|}{} & \multicolumn{2}{|c|}{{\bf 28}} & {\bf 27} & {\bf 27} \\ \cline{5-8}
\multicolumn{4}{|c|}{{\bf 28}} & {\bf 28} & {\bf 27} & {\bf 1} & {\bf 27} \\ \cline{5-8}
\multicolumn{4}{|c|}{} & {\bf 27} & {\bf 27} & {\bf 27} & {\bf 27} \\ \hline
\multicolumn{2}{|c|}{} & {\bf 28} & {\bf 27} & \multicolumn{2}{|c|}{} & {\bf 6} & {\bf 27} \\ \cline{3-4} \cline{7-8}
\multicolumn{2}{|c|}{{\bf 28}} & {\bf 27} & {\bf 27} & \multicolumn{2}{|c|}{{\bf 6}} & {\bf 27} & {\bf 27} \\ \hline
{\bf 28} & {\bf 27} & {\bf 1} & {\bf 27} & {\bf 6} & {\bf 27} & {\bf 1} & {\bf 27} \\ \hline
{\bf 27} & {\bf 27} & {\bf 27} & {\bf 27} & {\bf 27} & {\bf 27} & {\bf 27} & {\bf 27} \\ \hline
\end{tabular}
}
\\
\subtable[$\pmb{k=4}$]{
\setlength{\tabcolsep}{0.4mm}
\tiny
\begin{tabular}{|c|c|c|c|c|c|c|c|c|c|c|c|c|c|c|c|} \hline
\multicolumn{8}{|c|}{} &
\multicolumn{4}{|c|}{} & \multicolumn{2}{|c|}{} & {\bf 120} & \colorbox{lightgray}{\makebox[2em]{\bf 81}} \\ \cline{15-16}
\multicolumn{8}{|c|}{} &
\multicolumn{4}{|c|}{} & \multicolumn{2}{|c|}{{\bf 120}} & \colorbox{lightgray}{\makebox[2em]{\bf 81}} & \colorbox{lightgray}{\makebox[2em]{\bf 81}} \\ \cline{13-16}
\multicolumn{8}{|c|}{} &
\multicolumn{4}{|c|}{{\bf 120}} & {\bf 120} & \colorbox{lightgray}{\makebox[2em]{\bf 81}} & {\bf 1} & \colorbox{lightgray}{\makebox[2em]{\bf 81}} \\ \cline{13-16}
\multicolumn{8}{|c|}{{\bf\large 120}} &
\multicolumn{4}{|c|}{} & \colorbox{lightgray}{\makebox[2em]{\bf 81}} & \colorbox{lightgray}{\makebox[2em]{\bf 81}} & \colorbox{lightgray}{\makebox[2em]{\bf 81}} & \colorbox{lightgray}{\makebox[2em]{\bf 81}} \\ \cline{9-16}
\multicolumn{8}{|c|}{} &
\multicolumn{2}{|c|}{} & {\bf 120} & \colorbox{lightgray}{\makebox[2em]{\bf 81}} & \multicolumn{2}{|c|}{} & {\bf 6} & \colorbox{lightgray}{\makebox[2em]{\bf 81}} \\ \cline{11-12} \cline{15-16}
\multicolumn{8}{|c|}{} &
\multicolumn{2}{|c|}{{\bf 120}} & \colorbox{lightgray}{\makebox[2em]{\bf 81}} & \colorbox{lightgray}{\makebox[2em]{\bf 81}} & \multicolumn{2}{|c|}{{\bf 6}} & \colorbox{lightgray}{\makebox[2em]{\bf 81}} & \colorbox{lightgray}{\makebox[2em]{\bf 81}} \\ \cline{9-16}
\multicolumn{8}{|c|}{} &
{\bf 120} & \colorbox{lightgray}{\makebox[2em]{\bf 81}} & {\bf 1} & \colorbox{lightgray}{\makebox[2em]{\bf 81}} & {\bf 6} & \colorbox{lightgray}{\makebox[2em]{\bf 81}} & {\bf 1} & \colorbox{lightgray}{\makebox[2em]{\bf 81}} \\ \cline{9-16}
\multicolumn{8}{|c|}{} &
\colorbox{lightgray}{\makebox[2em]{\bf 81}} & \colorbox{lightgray}{\makebox[2em]{\bf 81}} & \colorbox{lightgray}{\makebox[2em]{\bf 81}} & \colorbox{lightgray}{\makebox[2em]{\bf 81}} & \colorbox{lightgray}{\makebox[2em]{\bf 81}} & \colorbox{lightgray}{\makebox[2em]{\bf 81}} & \colorbox{lightgray}{\makebox[2em]{\bf 81}} & \colorbox{lightgray}{\makebox[2em]{\bf 81}} \\ \hline
\multicolumn{4}{|c|}{} & \multicolumn{2}{|c|}{} & {\bf 120} & \colorbox{lightgray}{\makebox[2em]{\bf 81}} &
   \multicolumn{4}{|c|}{} & \multicolumn{2}{|c|}{} & {\bf 28} & \colorbox{lightgray}{\makebox[2em]{\bf 81}}
\\ \cline{7-8} \cline{15-16}
\multicolumn{4}{|c|}{} & \multicolumn{2}{|c|}{{\bf 120}} & \colorbox{lightgray}{\makebox[2em]{\bf 81}} & \colorbox{lightgray}{\makebox[2em]{\bf 81}} &
   \multicolumn{4}{|c|}{} & \multicolumn{2}{|c|}{{\bf 28}} & \colorbox{lightgray}{\makebox[2em]{\bf 81}} & \colorbox{lightgray}{\makebox[2em]{\bf 81}}
\\ \cline{5-8} \cline{13-16}
\multicolumn{4}{|c|}{{\bf 120}} & {\bf 120} & \colorbox{lightgray}{\makebox[2em]{\bf 81}} & {\bf 1} & \colorbox{lightgray}{\makebox[2em]{\bf 81}} &
   \multicolumn{4}{|c|}{{\bf 28}} & {\bf 28} & \colorbox{lightgray}{\makebox[2em]{\bf 81}} & {\bf 1} & \colorbox{lightgray}{\makebox[2em]{\bf 81}}
\\ \cline{5-8} \cline{13-16}
\multicolumn{4}{|c|}{} & \colorbox{lightgray}{\makebox[2em]{\bf 81}} & \colorbox{lightgray}{\makebox[2em]{\bf 81}} & \colorbox{lightgray}{\makebox[2em]{\bf 81}} & \colorbox{lightgray}{\makebox[2em]{\bf 81}} &
   \multicolumn{4}{|c|}{} & \colorbox{lightgray}{\makebox[2em]{\bf 81}} & \colorbox{lightgray}{\makebox[2em]{\bf 81}} & \colorbox{lightgray}{\makebox[2em]{\bf 81}} & \colorbox{lightgray}{\makebox[2em]{\bf 81}}
\\ \hline
\multicolumn{2}{|c|}{} & {\bf 120} & \colorbox{lightgray}{\makebox[2em]{\bf 81}} & \multicolumn{2}{|c|}{} & {\bf 6} & \colorbox{lightgray}{\makebox[2em]{\bf 81}} &
   \multicolumn{2}{|c|}{} & {\bf 28} & \colorbox{lightgray}{\makebox[2em]{\bf 81}} & \multicolumn{2}{|c|}{} & {\bf 6} & \colorbox{lightgray}{\makebox[2em]{\bf 81}}
\\ \cline{3-4} \cline{7-8} \cline{11-12} \cline{15-16}
\multicolumn{2}{|c|}{{\bf 120}} & \colorbox{lightgray}{\makebox[2em]{\bf 81}} & \colorbox{lightgray}{\makebox[2em]{\bf 81}} & \multicolumn{2}{|c|}{{\bf 6}} & \colorbox{lightgray}{\makebox[2em]{\bf 81}} & \colorbox{lightgray}{\makebox[2em]{\bf 81}} &
   \multicolumn{2}{|c|}{{\bf 28}} & \colorbox{lightgray}{\makebox[2em]{\bf 81}} & \colorbox{lightgray}{\makebox[2em]{\bf 81}} & \multicolumn{2}{|c|}{{\bf 6}} & \colorbox{lightgray}{\makebox[2em]{\bf 81}} & \colorbox{lightgray}{\makebox[2em]{\bf 81}}
\\ \hline
{\bf 120} & \colorbox{lightgray}{\makebox[2em]{\bf 81}} & {\bf 1} & \colorbox{lightgray}{\makebox[2em]{\bf 81}} & {\bf 6} & \colorbox{lightgray}{\makebox[2em]{\bf 81}} & {\bf 1} & \colorbox{lightgray}{\makebox[2em]{\bf 81}} &
   {\bf 28} & \colorbox{lightgray}{\makebox[2em]{\bf 81}} & {\bf 1} & \colorbox{lightgray}{\makebox[2em]{\bf 81}} & {\bf 6} & \colorbox{lightgray}{\makebox[2em]{\bf 81}} & {\bf 1} & \colorbox{lightgray}{\makebox[2em]{\bf 81}}
\\ \hline
\colorbox{lightgray}{\makebox[2em]{\bf 81}} & \colorbox{lightgray}{\makebox[2em]{\bf 81}} & \colorbox{lightgray}{\makebox[2em]{\bf 81}} & \colorbox{lightgray}{\makebox[2em]{\bf 81}} & \colorbox{lightgray}{\makebox[2em]{\bf 81}} & \colorbox{lightgray}{\makebox[2em]{\bf 81}} & \colorbox{lightgray}{\makebox[2em]{\bf 81}} & \colorbox{lightgray}{\makebox[2em]{\bf 81}} &
   \colorbox{lightgray}{\makebox[2em]{\bf 81}} & \colorbox{lightgray}{\makebox[2em]{\bf 81}} & \colorbox{lightgray}{\makebox[2em]{\bf 81}} & \colorbox{lightgray}{\makebox[2em]{\bf 81}} & \colorbox{lightgray}{\makebox[2em]{\bf 81}} & \colorbox{lightgray}{\makebox[2em]{\bf 81}} & \colorbox{lightgray}{\makebox[2em]{\bf 81}} & \colorbox{lightgray}{\makebox[2em]{\bf 81}}
\\ \hline
\end{tabular}
}
\caption{\label{f3}Contribution to \PAR\ for $k=0,1,2,3,4$.}
\end{figure}

Based on these observations, we can obtain a recurrence for the total contribution to the average
\PAR\ of all the tiles in a generic matrix.
We need the following parameters:
\begin{itemize}
\item
The number of bits in the original matrix, that we denote by $k$;

\item
The number of bits corresponding to the size of the matrix, or submatrix being considered,
that we denote by $i$;

\item
The number of bits to be used in the calculation of the contribution of the upper-left portion of the matrix, or submatrix, being considered;
we denote this by $j$.
\end{itemize}
The recurrence that computes the total contribution to the PAR of all the tiles in the matrix is:
\[
g\left(i,j,k\right)=\left\{\begin{array}{ll}
3^k, & \mbox{if $i=0$} \\
2^{2j-1}-2^{j-1}+2g\left(i-1,j,k\right)+g\left(i-1,i-1,k\right), & \mbox{otherwise} \\
\end{array}
\right.
\]
The values of $i$ and $j$ are initially set to the value of $k$.
The interpretation of each term in the above recurrence is as follows:
\begin{itemize}
\item
$3^k$ is the contribution of each ``background tile'';

\item
$2^{2j-1}-2^{j-1}$ is the contribution of the first quadrant;

\item
$g\left(i-1,j,k\right)$ is the contribution of each one of the second and third quadrants and

\item
$g\left(i-1,i-1,k\right)$ is the contribution of the fourth quadrant.
\end{itemize}
Remember that, for a given $k$, the recurrence equation is initialized with $i=j=k$. Thus, we have:
\begin{description}
\item[Case: $\mathbf{k=0}$:] $g\left(k,k,k\right)=3^k=3^{2k}$.

\item[Case: $\mathbf{k>0}$:]
$g(k,k,k)=g(k-1,k-1,k)+2g(k-1,k,k)+t(k)$.
The second parameter to the function indicates how to generate the $t(k)$ terms; the value of such terms is proportional to that parameter.
Thus, for $a\geq b$, $g(k,a,k)\ge g(k,b,k)$.
For our lower bound, we can neglect the terms $t(k)$. Thus, we obtain:
\begin{multline*}
g(k,k,k)\ge 3g(k-1,k-1,k)\ge 3g(k-2,k-2,k)
\\
\ge\cdots\cdots\ge 3g(1,1,k)\ge 3g(0,0,k)
\end{multline*}
For each step, the value of the first parameter decreased exactly by one unit, so after $k$ iterations the value of the
first parameter will be zero. Hence we have
$g(k,k,k)\ge 3^kg(0,0,k)$.
Since $g(0,0,k)=3^k$
we finally obtain
$g(k,k,k)\ge 3^k\times 3^k=3^{2k}$.
\end{description}
Thus, $\alpha_{\scriptscriptstyle\UD}=g(k,k,k)/2^{2k}\ge {\left(3/2\right)}^{2k}$.
\end{proof}

\subsection{Equality function}

\begin{Theorem}
$\alpha_{\scriptscriptstyle\UD}=2^k-2+2^{1-k}$
and
$\aw=2^{2k-1}-2^{k-1}$.
\end{Theorem}

\begin{proof}
An illustration of the ideal partition into monochromatic {\em regions}
for equality function is shown in Fig.~\ref{eq1}{\bf (a)}.
After running the bisection protocol, the induced tiling is (for $k=3$)
is shown in Fig.~\ref{eq1}{\bf (b)}.
Excluding the diagonal, we have $2$ tiles of size $16$, $4$ tiles of size $4$, and $8$ tiles of size $1$.
In general, it is easy to observe that, for each $0\leq i<k$, we have exactly $2^{k-i}$ tiles of size $2^{2i}$.

\begin{figure}[htbp]
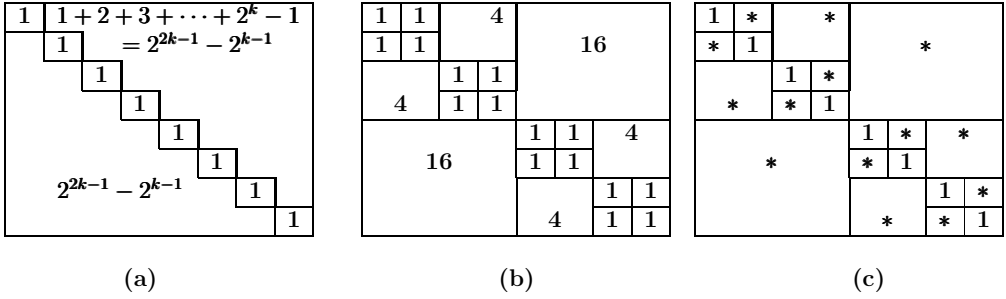

\scalebox{0.76}
{
\begin{tabular}{|c|c|c|c|c|c|c|c|cc|c|c|c|c|c|c|c|c|c|cc|c|c|c|c|c|c|c|c|c|c|} \cline{1-8} \cline{11-18} \cline{21-28}
{\bf 1} & \multicolumn{7}{|c|}{$\pmb{1+2+3+\dots+2^k-1}$}
  & & &
   {\bf 1}  & {\bf 1}  & \multicolumn{2}{|r|}{\bf 4} & \multicolumn{4}{|c|}{}
     & \multicolumn{2}{c|}{} &
      {\bf 1}  & $\pmb{\ast}$    & \multicolumn{2}{|r|}{$\pmb{\ast}$} & \multicolumn{4}{|c|}{}
\\ \cline{1-2} \cline{11-12} \cline{21-22}
\multicolumn{1}{|c|}{} & {\bf 1} & \multicolumn{6}{|c|}{$\pmb{=2^{2k-1}-2^{k-1}}$}
  & & &
   {\bf 1}  & {\bf 1}  & \multicolumn{2}{|c|}{} & \multicolumn{4}{|c|}{\bf 16}
     & \multicolumn{2}{c|}{} &
         $\pmb{\ast}$  & {\bf 1}  & \multicolumn{2}{|c|}{} & \multicolumn{4}{|c|}{$\pmb{\ast}$}
\\ \cline{2-3} \cline{11-14} \cline{21-24}
\multicolumn{2}{|c|}{} & {\bf 1} & \multicolumn{5}{|c|}{}
  & & &
    \multicolumn{2}{c|}{} & {\bf 1}  & {\bf 1}  & \multicolumn{4}{|c|}{}
      & \multicolumn{2}{c|}{} &
    \multicolumn{2}{c|}{} & {\bf 1}  & $\pmb{\ast}$  & \multicolumn{4}{|c|}{}
\\ \cline{3-4} \cline{13-14} \cline{23-24}
\multicolumn{3}{|c|}{} & {\bf 1} & \multicolumn{4}{|c|}{}
  & & &
    \multicolumn{2}{c|}{\bf 4} & {\bf 1}  & {\bf 1}  & \multicolumn{4}{c|}{}
      & \multicolumn{2}{c|}{} &
        \multicolumn{2}{c|}{$\pmb{\ast}$} & $\pmb{\ast}$  & {\bf 1}  & \multicolumn{4}{c|}{}
\\ \cline{4-5} \cline{11-18} \cline{21-28}
\multicolumn{4}{|c|}{} & {\bf 1} & \multicolumn{3}{|c|}{}
  & & &
    \multicolumn{4}{c|}{} & {\bf 1}  & {\bf 1}  & \multicolumn{2}{c|}{\bf 4}
      & \multicolumn{2}{c|}{} &
        \multicolumn{4}{c|}{} & {\bf 1}  & $\pmb{\ast}$  & \multicolumn{2}{c|}{$\pmb{\ast}$}
\\ \cline{5-6} \cline{15-16} \cline{25-26}
\multicolumn{5}{|c|}{} & {\bf 1} & \multicolumn{2}{|c|}{}
  & & &
    \multicolumn{4}{c|}{\bf 16} & {\bf 1}  & {\bf 1}  & \multicolumn{2}{c|}{}
      & \multicolumn{2}{c|}{} &
         \multicolumn{4}{c|}{$\pmb{\ast}$} & $\pmb{\ast}$  & {\bf 1}  & \multicolumn{2}{c|}{}
\\ \cline{6-7} \cline{15-18} \cline{25-28}
\multicolumn{6}{|c|}{$\pmb{2^{2k-1}-2^{k-1}}$} & {\bf 1} & \multicolumn{1}{|c|}{}
  & & &
    \multicolumn{4}{c}{} & \multicolumn{2}{|c|}{} & {\bf 1}  & {\bf 1}
      & \multicolumn{2}{c|}{} &
        \multicolumn{4}{c}{} & \multicolumn{2}{|c|}{} & {\bf 1}  &  $\pmb{\ast}$
\\ \cline{7-8} \cline{17-18} \cline{27-28}
\multicolumn{7}{|c|}{} & {\bf 1}
  & & &
    \multicolumn{4}{c}{} & \multicolumn{2}{|c|}{\bf 4} & {\bf 1}  & {\bf 1}
      & \multicolumn{2}{c|}{} &
        \multicolumn{4}{c}{} & \multicolumn{2}{|c|}{$\pmb{\ast}$} & $\pmb{\ast}$  & {\bf 1}
\\ \cline{1-8} \cline{11-18} \cline{21-28}
\multicolumn{27}{c}{}
\\
\multicolumn{7}{c}{\bf (a)}
 & \multicolumn{3}{c}{} &
   \multicolumn{8}{c}{\bf (b)}
      & \multicolumn{3}{c}{} &
        \multicolumn{7}{c}{\bf (c)}
\\
\end{tabular}
}
\caption{\label{eq1}{\bf (a)} Ideal tiling for equality function.
{\bf (b)} The induced tiling by the bisection protocol (shown for $k=3$).
{\bf (c)} Contribution of each rectangle in protocol-induced tiling
where $\pmb{\ast\,\equiv\,2^{2k-1}-2^{k-1}}$.
The numbers in the figure denote the size of each tile.}
\end{figure}

The following accounting scheme can be used to simplify calculation.
For uniform distribution $\UD$, $\alpha_{\scriptscriptstyle\UD}$ is
the sum of the ratio $\frac{|R^I(i,j)|}{|R^P(i,j)|}$ over each element $(i,j)$ in the matrix divided by the number of total elements $2^{2k}$ in the matrix,
where $R^I(i,j)$ and $R^P(i,j)$ is the size of the ideal and protocol-induced tiling that contains the cell $(i,j)$.
Consider a rectangle $A$ of size $m$ in the protocol-induced tiling
and suppose that $A$ is contained in a monochromatic region of the ideal partition of size $m'$.
Then, the sum of contributions of the elements of $A$ is
$\sum_{i=1}^m \frac{m'}{m}=m'$.
Thus, the total contribution of the rectangle $A$ is simply the size of region of the ideal
partition containing it.

Fig.~\ref{eq1}{\bf (c)} illustrates the contribution of each rectangle in the protocol-induced tiling to average \PAR.
We can calculate the total contribution to the average \PAR\ of all the tiles in the matrix, except the diagonal, by
multiplying $2^{2k-1}-2^{k-1}$ by the number of tiles.
The number of tiles is given by:
$\sum^{k-1}_{i=0}2^{k-i}=2^{k+1}-2$.
The total contribution of those tiles is
$(2^{k+1}-2)\times\left(2^{2k-1}-2^{k-1}\right)=2^{3k}-2^{2k+1}+2^k$.
The contribution of the diagonal is $\underbrace{1+1+\cdots\cdots+1}_{2^k\mbox{ \small times}}=2^k$.
Since the average objective PAR $\alpha_{\scriptscriptstyle\UD}$ is the sum of the total contributions divided by the number of cells in the matrix,
we have
\[
\alpha_{\scriptscriptstyle\UD}=\frac{2^{3k}-2^{2k+1}+2^k+2^k}{2^{2k}}=\frac{2^{3k}-2^{2k+1}+2^{k+1}}{2^{2k}}=
2^k-2+2^{1-k}
\]
It can be seen from the ideal and protocol tilings that the worst case for \PAR\ is the one in which the ideal tile size is
$2^{2k-1}-2^{k-1}$, and the protocol tile size is $1$.
Thus $\alpha_{\mathrm{worst}}=2^{2k-1}-2^{k-1}$.
\end{proof}


\begin{thebibliography}{99}
\bibitem{GRS09}
A. Ghosh, T. Roughgarden and M. Sundararajan.
{\em Universally utility-maximizing privacy mechanisms},
$41\tx$ ACM Symposium on Theory of Computing, 351-360, 2009.

\bibitem{BDM02}
P. Berman, B. DasGupta and S. Muthukrishnan.
{\em On the Exact Size of the Binary Space Partitioning of Sets of Isothetic Rectangles with Applications},
SIAM Journal of Discrete Mathematics, 15 (2), 252-267, 2002.

\bibitem{BCKO93}
R. Bar-Yehuda, B. Chor, E. Kushilevitz and A. Orlitsky.
{\em Privacy, additional information, and communication},
IEEE Transactions on Information Theory, 39, 55-65, 1993.

\bibitem{dAF92}
F. d'Amore and P. G. Franciosa.
{\em On the optimal binary plane partition for sets of isothetic rectangles},
Information Processing Letters, 44, 255-259, 1992.

\bibitem{CCD88}
D. Chaum, C. Cr\'{e}peau and I. Damgaard.
{\em Multiparty, unconditionally secure protocols},
$22^{\mathrm{th}}$ ACM Symposium on Theory of Computing, 11-19, 1988.

\bibitem{CK91}
B. Chor and E. Kushilevitz.
{\em A zero-one law for boolean privacy},
SIAM Journal of Discrete Mathematics, 4, 36-47, 1991.

\bibitem{D06}
C. Dwork.  {\em Differential privacy},
$33^{\mathrm{rd}}$ International Colloquium on Automata, Languages and Programming, 1-12, 2006.

\bibitem{FJS10}
J. Feigenbaum, A. Jaggard and M. Schapira.
{\em Approximate Privacy: Foundations and Quantification},
ACM Conference on Electronic Commerce, 167-178, 2010.

\bibitem{muller}
E. Grigorievaa, P. J.-J. Heringsb, R. M\"{u}llera and D. Vermeulena.
{\em The communication complexity of private value single-item auctions},
Operations Research Letters, 34, 491-498, 2006.

\bibitem{muller2}
E. Grigorievaa, P. J.-J. Heringsb, R. M\"{u}llera and D. Vermeulena.
{\em The private value single item bisection auction},
Economic Theory, 30, 107-118, 2007.

\bibitem{HN97}
E. Kushilevitz and N. Nisan. {\em Communication Complexity}, Cambridge University Press, 1997.

\bibitem{KL11}
D. Kifer and B.-R. Lin.
{\em An Axiomatic View of Statistical Privacy and Utility},
to appear in Journal of Privacy and Confidentiality (conference
version appeared in 2010 ACM SIGMOD/PODS Conference).

\bibitem{K92}
E. Kushilevitz.
{\em Privacy and communication complexity},
SIAM Journal of Discrete Mathematics, 5 (2), 273-284, 1992.

\bibitem{paterson90}
M. Paterson and F. F. Yao.
{\em Efficient binary space partitions for hidden-surface removal and solid modeling},
Discrete and Computational Geometry, 5(1), 485-503, 1990.

\bibitem{paterson92}
M. Paterson and F. F. Yao.
{\em Optimal binary space partitions for orthogonal objects},
Journal of Algorithms, 13, 99-113, 1992.

\bibitem{T05}
C. D. T\'{o}th.
{\em Binary Space Partitions: Recent Developments},
in Combinatorial and Computational Geometry,
J. E. Goodman, J. Pach and E. Welzl (eds.),
MSRI Publications, 52, 529-556, Cambridge University Press, 2005.

\bibitem{Y79}
A. C. Yao.
{\em Some complexity questions related to distributive computing},
$11^{\mathrm{th}}$ ACM Symposium on Theory of Computing, 209-213, 1979.
\end{thebibliography}
\end{document}